\documentclass[journal,comsoc]{IEEEtran}

\usepackage{latexsym}
\usepackage{epsfig}
\usepackage{color}
\usepackage{amsmath} 
\usepackage{amssymb}
\usepackage{amsxtra}
\usepackage{enumitem}

\usepackage[T1]{fontenc}
\usepackage{setspace}

\usepackage{amsmath}
\usepackage[cmintegrals]{newtxmath}
\usepackage{graphicx}
\graphicspath{{Figures/}}
\usepackage{url}
\usepackage{cite}
\usepackage{algorithm}
\usepackage{algpseudocode}
\usepackage{subcaption}

\makeatletter
\def\BState{\State\hskip-\ALG@thistlm}
\makeatother

\usepackage{epstopdf}
\usepackage{amsmath}
\usepackage{amsxtra}
\usepackage{amstext}
\usepackage{amssymb}
\usepackage{latexsym}
\usepackage{dsfont} 
\usepackage{color}
\usepackage{multirow}
\usepackage{float}
\usepackage{hyperref}

\usepackage{tabularx,colortbl}
\usepackage[table]{xcolor}

\usepackage{lscape}

\usepackage{algorithm}
\usepackage{algpseudocode}

\usepackage{cuted}
\usepackage{units}
\usepackage{cases}
\usepackage{mathtools}
\mathtoolsset{showonlyrefs} 

\newcommand{\mbf}[1]{\mathbf{#1}}

\newcommand{\nth}[1]{{#1}{\text{th}}}

\newcommand{\abs}[1]{\left|{#1}\right|}

\newcommand{\red}[1]{{\color{red}{#1}}} 

\usepackage{mathtools}

\newcommand{\Hr}{\mathcal{H}}
\newcommand{\spp}{\mathrm{spp}}
\newcommand{\LoS}{\mathrm{LoS}}
\newcommand{\NLoS}{\mathrm{NLoS}}

\newcommand{\opt}{\mathrm{opt}}
\newcommand{\SPP}{\mathrm{SPP}}
\newcommand{\maxx}{\mathrm{max}}
\newcommand{\T}{\mathrm{T}}

\newcommand{\clu}{\mathrm{clu}}
\newcommand{\ray}{\mathrm{ray}}
\newcommand{\noise}{\mathrm{noise}}
\newcommand{\sys}{\mathrm{sys}}
\newcommand{\mol}{\mathrm{mol}}

\begin{document}

\title{An Overview of Signal Processing Techniques for Terahertz Communications
}

\author{Hadi~Sarieddeen,~\IEEEmembership{Member,~IEEE,}
        Mohamed-Slim~Alouini,~\IEEEmembership{Fellow,~IEEE,}
        \\and~Tareq~Y.~Al-Naffouri,~\IEEEmembership{Senior Member,~IEEE}
\thanks{The authors are with the Department of Computer, Electrical and Mathematical Sciences and Engineering (CEMSE), King Abdullah University of Science and Technology (KAUST), Thuwal, Makkah Province, Kingdom of Saudi Arabia, 23955-6900 (e-mail: hadi.sarieddeen@kaust.edu.sa; slim.alouini@kaust.edu.sa; tareq.alnaffouri@kaust.edu.sa). This work was supported by the KAUST Office of Sponsored Research.
}
}
%
\maketitle

\begin{abstract}

Terahertz (THz)-band communications are a key enabler for future-generation wireless communications systems that promise to integrate a wide range of data-demanding applications. Recent advances in photonic, electronic, and plasmonic technologies are closing the gap in THz transceiver design. Consequently, prospect THz signal generation, modulation, and radiation methods are converging, and corresponding channel model, noise, and hardware-impairment notions are emerging. Such progress establishes a foundation for well-grounded research into THz-specific signal processing techniques for wireless communications. This tutorial overviews these techniques, emphasizing ultra-massive multiple-input multiple-output (UM-MIMO) systems and reconfigurable intelligent surfaces, vital for overcoming the distance problem at very high frequencies.  We focus on the classical problems of waveform design and modulation, beamforming and precoding, index modulation, channel estimation, channel coding, and data detection. We also motivate signal processing techniques for THz sensing and localization.  

\end{abstract}


\section{Introduction}
\label{sec:introduction}


\subsection{THz Communications are Emerging}

The wireless communication's frequency spectrum has been continuously expanding to satisfy ever-increasing bandwidth demands. Although millimeter-wave (mmWave)-band communications \cite{Xiao7959169,Rangan6732923} are already shaping the fifth-generation (5G) of wireless mobile communications, terahertz (THz)-band communications \cite{Akyildiz6882305,akyildiz2014terahertz,kleine2011review,Chen8663550,Huang5764977,Song6005345} will be essential to the future sixth-generation (6G) \cite{rajatheva2020white,rajatheva2020scoring,dang2020should,Huq8782882,Zhang8663549,boulogeorgos2018wireless,tariq2019speculative,Giordani9040264,han2019terahertz,dore2020technology,mourad2020baseline,saad2019vision,sarieddeen2019generation} and beyond. The THz band is sandwiched between the microwave and optical bands as the last unexplored area of the radio-frequency (RF) spectrum. Hence, technologies from both sides are being explored to support THz communications. RF engineers label as THz all operations beyond the 100 gigahertz (GHz) threshold, below which most known mmWave use cases exist. In contrast, optical engineers label as THz any frequency below $\unit[10]{THz}$ (the far-infrared). However, the THz range is $\unit[300]{GHz}$ to $\unit[10]{THz}$ according to IEEE \emph{Transactions on Terahertz Science and Technology} and closely mapped to the tremendously high frequency (THF) band  ($\unit[300]{GHz}$ to $\unit[3]{THz}$) according to ITU-R. 


THz communications are called-for despite the maturity of neighboring technologies. In contrast to mmWave communications, THz communications can exploit the available spectrum to achieve a terabits per second (Tbps) data rate without additional spectral efficiency enhancement techniques. {Furthermore, due to the shorter wavelengths, THz signals are less susceptible to free-space diffraction and inter-antenna interference and exhibit higher resilience to eavesdropping. THz systems can also support higher link directionality and can be achieved in much smaller footprints.} In contrast, {compared to visible light communications (VLC) \cite{Grobe6685758,Pathak7239528}, THz signals are not as severely affected by alignment issues}, ambient light, atmospheric turbulence, scintillation, fog, and temporary spatial variation in light intensity. THz communications can thus complement both mmWave and VLC by providing alternative quasi-optical paths. However, due to significant water vapor absorption above $\unit[1]{THz}$, a gap might always exist for wireless communications at the high end of the THz range.

Consequently, THz communications will enable ultra-high bandwidth THz communications are thus expected to enable ultra-high bandwidth, and ultra-low latency communication paradigms \cite{Rappaport8732419}. For example, they can be used to achieve optical-fiber-like performance in network backhauling \cite{han2019terahertz}, backbone (rack-to-rack) connectivity in data centers \cite{celik2018wireless,Boujnah8801998,Cheng9039668}, and high data rate kiosk-to-mobile communications \cite{He7928447}. Furthermore, THz wireless bridges enable the transparent integration of fiber networks without requiring detection, decoding, and re-modulation \cite{castro2020100}. THz links supporting 100 gigabits per second (Gbps) have already been demonstrated over distances corresponding to such applications \cite{nagatsuma2016advances}. However, the holy grail of THz communications is enabling mobile communications at the device level and, in the context of medium-range indoor, vehicular, drone-to-drone \cite{Xu9370130,Xia9153882}, or device-to-device communications, at the access level. When combined with other THz-band applications such as accurate localization, sensing, and imaging, THz communications can enable wireless remoting of human cognition, leading to ubiquitous wireless intelligence \cite{Rappaport8732419,Matti_Oulu}.


\subsection{Advances in THz Devices}
\label{Sec:intro_devices}


{

The main contributions to THz technology are still at the device level rather than the system level. High-frequency electromagnetic radiation is perceived either as waves processed via electronic devices (the mmWave realm) or as particles processed via photonic devices (the optical realm). In between, the THz band is dubbed as a ``THz gap'' due to the lack of compact THz signal sources and detectors that have high power and sensitivity, respectively. However, recent electronic and photonic THz transceiver design advances have enabled efficient signal generation, modulation, and radiation \cite{Kenneth8808165,nagatsuma2016advances,sengupta2018terahertz,koenig2013wireless,Sengupta9118734}. 

In electronic solutions \cite{hillger2020toward,rieh2020introduction}, silicon-based devices \cite{Hillger8576551, Han8661721} successfully used for mmWave systems are being advocated for the THz range. Electronic solutions are based primarily on silicon complementary metal-oxide-semiconductor (CMOS) and silicon-germanium (SiGe) BiCMOS technologies \cite{Nikpaik8077757,Aghasi7819530,mittleman2017perspective,Heinemann7838335}, and have demonstrated incredible compactness and compatibility with existing fabrication processes. CMOS solutions exhibit lower speed and power handling capabilities, but are the most compact. However, the corresponding highest unity current gain frequency ($f_{\T}$) and unity maximum available power gain frequency ($f_{\maxx}$) remain at $\unit[200]{GHz}$ and $\unit[320]{GHz}$, respectively. Nevertheless, higher operating frequencies have been noted by III-V-based (Indium phosphide (InP)-based \cite{Urteaga7915698}, for example) semiconductors \cite{Sengupta8757165} in high electron mobility transistors (HEMTs) \cite{Deal8240460,Leuther6997797,Mei7047678}, heterojunction bipolar transistors (HBTs) \cite{Urteaga7915698,Bolognesi7838506}, and Schottky diodes \cite{Mehdi7835091}. The $f_{\maxx}$ of III-V based devices have crossed $\unit[1.5]{THz}$ \cite{Sengupta9118734}, and in \cite{Heinemann7838335}, SiGe-HBT devices could operate at $\unit[720]{GHz}$. Resonant tunneling diodes (RTDs) have also exhibited oscillation frequencies up to $\unit[2]{THz}$ \cite{Feiginov10106313667191, Koyama2013, Maekawa2016, Izumi8066877}. Other demonstrations at different frequencies and powers are highlighted in \cite{Ahmed8100752, Hillger7969042, Han7265095, Hu8254345}.

In photonic solutions \cite{nagatsuma2016advances}, where the main design driver is the data rate, higher carrier frequencies are supported, but the degrees of integration and output power remain low (photonic devices have a larger form factor and are more expensive). Frequencies beyond $\unit[300]{GHz}$ have been supported using optical downconversion systems \cite{nagatsuma2016advances}, quantum cascade lasers \cite{lu2016room}, photoconductive antennas \cite{huang2017globally}, and uni-traveling carrier photodiodes \cite{Song6213156}. Among other advances in THz photonics, knowledge on the topological phase of light is being exploited to demonstrate robust THz topological valley transport through several sharp bends in on-chip THz waveguides \cite{yang2020terahertz}.



Satisfying emerging system-level properties requires designing efficient and programmable devices. This deviation from designing perfect THz devices resulted in the emergence of integrated hybrid electronic-photonic systems \cite{sengupta2018terahertz}, such as combining photonic transmitters and III-V electronic receivers. However, more precise synchronization between the transmitter and receiver is required in hybrid solutions. Integrating miniature micro-electro-mechanical systems (mMEMS) within THz components also promises high degrees of reconfigurability and enhanced performance. By enabling electronic control of frequency, polarization, and beam steering, mMEMS can achieve features beyond what can be achieved in state-of-the-art optical and semiconductor technologies.

Also gaining popularity are plasmonic solutions \cite{Jornet5995306,hafez2018extremely}, where novel plasmonic materials such as graphene exhibit high electron mobility and reconfigurability \cite{ju2011graphene,novoselov2012roadmap,ferrari2015science,Xu6846379}. The resultant surface plasmon polariton (SPP) waves in plasmonic antennas have much smaller resonant wavelengths than free space waves, which results in compact and flexible antenna array designs \cite{Singh9110885}. By leveraging the properties of plasmonic nanomaterials and nanostructures, transceivers and antennas that intrinsically operate at THz frequencies can be created, avoiding the upconversion and downconversion losses of electronic and photonic systems, respectively. Graphene can be used to develop direct THz signal sources, modulators (that manipulate amplitude, frequency, and phase), and on-chip THz antenna arrays \cite{thawdar2018modeling}. Graphene solutions promise to have low electronic noise temperature and generate energy-efficient short pulses \cite{Jornet6804405}. Alongside miniaturization, graphene nanoantennas also promise high directivity and radiation efficiency \cite{dash2020nanoantennas}. All these advances demonstrate that the gaps germane to designing THz technology are rapidly closing and that the THz-band will soon open for everyday applications.

Power consumption remains a significant hurdle toward the practical deployment of THz systems. Although studies on the power consumption of mmWave and sub-THz receivers are emerging, as reported in \cite{skrimponis2020power,faisal2019ultra}, the power consumption of true-THz devices is still not well-established. For fully digital architectures, \cite{skrimponis2020power} proposes design options for mixers, noise amplifiers, local oscillators, and analog-to-digital converters (ADCs) by assuming a $\unit[90]{nm}$ SiGe BiCMOS technology. The corresponding optimizations can reduce power by 80\% (fully-digital $\unit[140]{GHz}$ receiver with a $\unit[2]{GHz}$ sampling rate and power consumption less than $\unit[2]{W}$). 
Given that cellular systems do not usually require high output signal-to-noise-ratio (SNRs), the researchers propose relaxing linearity requirements and reducing the power consumption of components. Low-resolution ADCs and mixer power consumption are significantly reduced, and fully digital architectures introduce significant benefits compared to analog beamforming with comparable power consumption under bit-width optimization. For a hybrid array-of-subarrays (AoSA) multiple-input multiple-output (MIMO) structure at $\unit[28]{GHz}$ and $\unit[140]{GHz}$, and assuming a single-carrier (SC) system, the power dissipated by two major power-consuming components, namely the low noise amplifier (LNA) and the ADC, is estimated in \cite{faisal2019ultra}. The power consumption of a $\unit[140]{GHz}$ receiver is much higher than that of a $\unit[28]{GHz}$ receiver, indicating that considerable measures need to be taken to improve power consumption in high-frequency devices.

}

{
\subsection{Standardization Efforts}

Various leading 6G initiatives investigate THz communications, including the 6Genesis Flagship Program (6GFP) in Finland and the European Commission's Horizon 2020 ICT-09 THz Project Cluster. The U.S. Defense Advanced Research Projects Agency (DARPA) identifies THz technology as one of four major research areas that could impact society more than the Internet. Accordingly, THz-related research has attracted significant funding, and standardization efforts have been launched \cite{kurner2014towards,tekbiyik2019terahertz,elayan2019terahertz,tripathi2021millimeter,tripathi2021millimeter}. The Federal Communications Commission (FCC) has recently allocated \unit[21.2]{GHz} of spectrum between the $\unit[116]{GHz}$ and $\unit[246]{GHz}$ bands for unlicensed usage. Furthermore, the first IEEE 802.15.3d standardization efforts for sub-THz communications towards 6G were reported in \cite{Petrov9269931}, focusing on point-to-point links that can support 100 Gbps over few centimeters to several hundreds of meters. 

The first IEEE standard, 802.15.3d, was approved in 2017, with channels in the sub-THz range of 253-322 GHz and bandwidths up to 69 GHz (starting from 2.16 GHz), featuring up to 69 overlapping channels between 252.72 GHz and 321.84 GHz. All these bands were cleared for THz use at the World Radiocommunication Conference (WRC) 2019 (160 GHz of spectrum), without specifying the necessary conditions to protect passive services such as the radio astronomy earth exploration-satellite service \cite{Kurner9166206}. However, the potential interference with existing ground-to-orbit links should not be overlooked. The standard includes medium access (MAC) and physical (PHY) layer designs. MAC considerations include initial solutions of directional channel access, neighbor discovery, and synchronization. {More relevant to our work}, the PHY layer considerations are in two modes, THz SC mode (THz-SC PHY) and THz on-off keying mode (THz-OOK PHY). THz-SC PHY supports high rates in fronthaul and backhaul links and data centers, where complex signals are used, including quadrature amplitude modulation (QAM) up to 64-QAM. THz-OOK PHY is more tailored to short-distance (up to several meters) kiosk and intra-device communications, featuring lower-cost sub-THz devices and a single low-complexity modulation scheme, on-off keying keying (OOK), that can still achieve tens of Gbps. Both high-rate (14/15) and low-rate (11/15) low-density parity-check (LDPC) codes are enabled in both modes, but with THz-OOK PHY, an additional Reed Solomon (RS) code with simple decoding without soft decision information is also enabled. The target minimal receiver sensitivity levels in both modes are set to –67 dBm (for 11/15 LDPC and 2.16 GHz bandwidth).

IEEE 802.15.3d-compliant solutions are already being proposed, as in \cite{shehata2021ieee}, with THz waveform designs that exploit 99.3\% of the total in-band energy and enable out-of-band interference management. In this paper, we expand on the tradeoffs between SC and multi-carrier designs to enhance future standards. We address signal processing solutions suitable for both modes of the standard, and we formulate signal processing problems that require further consideration in future THz standardization efforts. Because IEEE 802.15.3d is tailored to fixed point-to-point links, we define a generic system model that is quasi-deterministic and maintain the discussion at the link level, by assuming the transmit and receive antenna directions are known. We assume interference to be mitigated by appropriate link planning with simple procedures for initial access and device discovery. However, we also address the system-level considerations of interference, link mobility, and multiple-channel access. The stationarity in point-to-point links introduces significant complexity reductions, much needed for Tbps processing, as we highlight in the subsequent sections.

}

\subsection{Significance of Signal Processing for THz Communications}

Many signal processing and communications system challenges still need to be addressed. The factors to be considered in signal processing in the THz realm differ significantly from those in the systems at lower frequencies; they are closely linked to the transceiver or device architectures. Efficient THz-band signal processing is crucial for two reasons. First, signal processing must account for the use of ultra-massive MIMO (UM-MIMO) antenna systems \cite{akyildiz2016realizing,faisal2019ultra,Busari8761371} to overcome the very short communication distances due to severe power limitations and propagation losses. Second, signal processing must overcome the mismatch between the bandwidths of the THz channel and the digital baseband system \cite{corre2019sub,Weithoffer8109974}.

{Because channel coding} is the most computationally demanding component of the baseband chain, several projects are studying efficient coding schemes for Tbps operations \cite{Kestel8625324}. However, the complete chain should be efficient and parallelizable. Therefore, joint algorithm and architecture co-optimization of channel estimation, channel coding, and data detection is required. Furthermore, the inherent sparsity in the angle and delay domains at THz frequencies can be exploited in solutions based on compressive sensing techniques. Low-resolution digital-to-analog conversion systems can also reduce the baseband complexity; even all-analog THz solutions have been considered.

Although THz communications exhibit quasi-optical traits, they retain several microwave characteristics. They can still use UM-MIMO antenna array processing techniques to support efficient beamforming and reflective surfaces to support non-line-of-sight (NLoS) propagation. Thus, efficient beamforming and beamsteering techniques and low-complexity precoding and combining algorithms are required. However, what seems predicted to be the norm in future THz systems are hybrid and adaptive AoSA antenna architectures, in which each subarray (SA) undergoes independent beamforming. Furthermore, given the large degrees of freedom in THz UM-MIMO systems at the transmitter side, several probabilistic shaping and index modulation schemes can be explored. This is particularly true in plasmonic solutions where each antenna element (AE) can be turned on and off or tuned to a specific frequency by simple material doping or electrostatic bias.

Molecular absorptions also result in band splitting and spectrum shrinking at larger communication distances. Distance-adaptive solutions in which antenna array designs and resource allocation criteria are optimized {can tackle} spectrum shrinking \cite{zakrajsek2017design,Han7321055}. Accordingly, the classical problems of waveform design and modulation need to be revisited. For instance, SC modulations can be favored over orthogonal frequency-division multiplexing (OFDM), which is challenging to implement in the THz band. {Nevertheless, in some indoor }THz scenarios, several multipath components might persist, resulting in frequency-selective channels. Frequency-selectivity might also arise at the receiver side also due to the behavior of THz components. {Therefore, multi-carrier modulations might still be required, perhaps in the form of multiple orthogonal and independent SCs that can be achieved at low complexity and combined with a form of carrier aggregation.}



\begin{figure*}[t]
  \centering
  \includegraphics[width=0.94\textwidth]{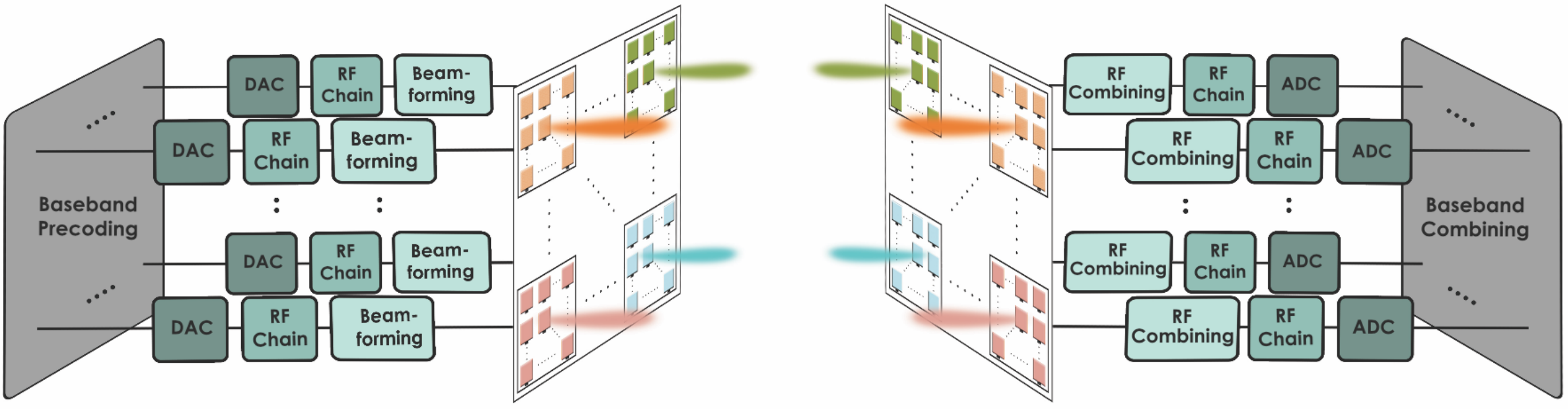}
 \caption{Typical THz-band communications system model.}
  \label{Fig_RF}
\end{figure*}



\begin{table}[t]
\footnotesize
\centering
\caption{{Relevant overview articles on THz communications.}}
\label{Tablerelatedsurveys}
\begin{tabular}{|p{2.5cm}|p{5.5cm}|}
\hline
\hline
 \textbf{Ref.}      & \textbf{Area of Focus}  \\ \hline\hline

Kurner \textit{et al.} \cite{kurner2014towards}  - 2014  & Focuses on THz channel modeling and standardization efforts.  \\ \hline

Akyildiz \textit{et al.} \cite{akyildiz2014terahertz}   - 2014   &  Focuses on THz applications at the nano and macro levels and presents possible THz band transceivers and antennas; also summarizes challenges in the physical layer.  \\ \hline

Nagatsuma \textit{et al.} \cite{nagatsuma2016advances,kleine2011review}  - 2016  & Overviews recent THz advances at the device level.  \\ \hline

Boulogeorgos \textit{et al.} \cite{Boulogeorgos8387218}  - 2018  & Highlightes how THz can deliver optical network quality; also touches on the channel, baseband challenges, and on beamforming. \\ \hline

Akyildiz \textit{et al.} \cite{Akyildiz8387211}  - 2018   & Highlights the importance of MIMO configurations and distance-adaptive solutions to extend the THz communication distance.  \\ \hline

Rappaport \textit{et al.} \cite{Rappaport8732419}  - 2019  & Details the opportunities and challenges in communications above 100 GHz towards 6G, mainly shaped by recent channel measurement campaigns led by the group. \\ \hline

Ghafoor \textit{et al.} \cite{ghafoor2019mac}   - 2019    & Presents THz-band applications and various MAC-layer protocols.  \\ \hline

Lemic \textit{et al.} \cite{lemic2019survey} - 2019   & Focuses on THz communications for nanonetworks and the corresponding applications and networking considerations, especially in in-body environments.  \\ \hline

Elayan \textit{et al.} \cite{elayan2019terahertz}   - 2020   &  Presents a timeline of THz research and funding and standardization activities, comparing THz to rival technologies. \\ \hline

Sarieddeen \textit{et al.} \cite{sarieddeen2019generation}  - 2020  & Highlights the importance of merging communications, sensing, imaging, and localization in the THz band.  \\ \hline

Tripathi \textit{et al.} \cite{tripathi2021millimeter}    - 2021   & Captures the physical layer differences between mmWave and THz communications. \\ \hline

Chaccour \textit{et al.} \cite{chaccour2021seven}  - 2021  & Describes joint THz communications and sensing, focusing on the possible applications and touching on both networking and physical layer considerations.  \\ \hline

\hline
\hline
\end{tabular}
\end{table}

{
\subsection{Paper Contributions and Outline}

In this paper, we provide an overview of recent advances in signal processing techniques for THz communications. It is unclear whether recent advances in THz devices have bridged the THz gap, primarily because there remains a gap between the promised data rates and the limited baseband-processing capabilities. Therefore, we seek to establish a clear link between novel THz devices, THz-channel and -noise models, and THz signal-processing techniques. Accordingly, we strive to formulate THz-specific signal processing methods that can alleviate the quasi-optical behavior of THz signals to provide seamless connectivity and accurate localization and sensing capacities. 

{Samples of popular recent surveys} and tutorials on THz communications in the literature are illustrated in Table \ref{Tablerelatedsurveys}, including notable papers written by pioneers in the field. The literature lacks a holistic approach to THz signal processing for communications and sensing, a nascent research field not well discovered but with a promising outlook. Papers by Rappaport \textit{et al.} are more tailored to channel measurement and modeling campaigns and channel sounders; those by Kurner \textit{et al.} are focused on channel modeling and standardization; those by Akyildiz \textit{et al.} discuss THz signal processing but not in detail; and those by Nagatsuma \textit{et al.} are focused on advances in THz devices.

We start our paper by examining how THz technology is widely believed to be the ultimate spectrum enabler for future-generation wireless communications systems. We then describe how the critical challenges in developing THz communications systems are, besides those inherent at the device level, at the infrastructure and algorithmic levels. We promote solutions on both levels, but most importantly, highlight the practical signal processing and hardware considerations critical to achieving the promised Tbps data rates, especially given the limitations in digital baseband solutions. The proposed signal-processing solutions should be low-cost and low-power-consuming. They should also be highly reliable, low-latency solutions. 

{Approaching THz research from }a signal-processing perspective is timely and should not wait until THz devices are mature. While we continue to monitor the latest THz devices and channel models to update our signal-processing solutions, this paper aims to study the signal-processing challenges and prospect THz use cases to guide research on THz transceivers. This paper's novel contributions are thus expected to catalyze THz research by directly linking its components to broadly and significantly enable future 6G wireless systems to achieve an order-of-magnitude increase in capacity with improved robustness.

The paper is organized as follows. The system model is first presented in Sec.~\ref{sec:sysmodel}, followed by a discussion on THz channel and noise modeling in Sec.~\ref{sec:channel_mod}. Then, recent performance analysis frameworks and experimental testbeds are summarized in Sec.~\ref{sec:analysis}. The latest advances in THz modulation and waveform designs are summarized in Sec.~\ref{sec:modulation}, the concept of THz ``spatial tuning'' and reconfigurable arrays are studied in Sec. \ref{sec:SPtuning}, and THz beamforming and precoding techniques are discussed in Sec.~\ref{sec:beamforming}. The baseband signal processing problems of THz channel estimation, channel coding, and data detection are then illustrated in Sec.~\ref{sec:baseband}. We examine the signal processing aspects of intelligent reflecting surface (IRS)-assisted THz communications in Sec.\ref{sec:surfaces}. Finally, in Sec.\ref{sec:sensing}, we highlight the importance of signal processing for THz sensing, imaging, and localization, briefly discussing THz networking and security before concluding in Sec. \ref{sec:conc}. Concerning notation, lower case, bold lower case, and bold upper case letters correspond to scalars, vectors, and matrices, respectively. We denote by $(\cdot)^{T}$, $(\cdot)^{\mathcal{H}}$, and $\mathsf{E}[\cdot]$, the transpose, conjugate transpose, and expected value, respectively. Table \ref{table:abre} summarizes the abbreviations.

}

\begin{table}[t]
\small
\caption{Frequently-Used Abbreviations}\label{table:abre}
\centering
 \begin{tabular}{|p{1.6cm} || p{6cm} |}
 \hline
3D & three-dimensional \\ \hline
5G & fifth-generation \\ \hline
6G & sixth-generation \\ \hline
ADC & analog-to-digital converter \\ \hline
AE & antenna element \\ \hline
AoA &  angle of arrival \\ \hline
AoD &  angle of departure \\ \hline
AoSA & array-of-subarrays \\ \hline
AWGN & additive white Gaussian noise \\ \hline
CFO & carrier frequency offset \\ \hline
CMOS & complementary metal-oxide-semiconductor\\ \hline
CPM & continuous phase modulation \\ \hline
CSI & channel state information \\ \hline
DAC & digital-to-analog converter \\ \hline
DAoSA & dynamic array-of-subarrays \\ \hline
Gbps & gigabit/second \\ \hline
GHz & gigahertz \\ \hline
GSM & generalized spatial modulation \\ \hline
GVD & group velocity dispersion \\ \hline
HITRAN & high-resolution transmission molecular absorption database \\ \hline
IM & index modulation \\ \hline
IRS & intelligent reflecting surface \\ \hline 
ISI & inter-symbol interference \\ \hline 
LDPC & low-density parity-check \\ \hline 
LIS & large intelligent surface \\ \hline 
LLR & log-likelihood-ratio \\ \hline 
LoS & line-of-sight \\ \hline
MIMO & multiple-input multiple-output \\ \hline
mmWave & millimeter-wave \\ \hline
NLoS & non-line-of-sight \\ \hline
NOMA & non-orthogonal multiple access \\ \hline
OFDM & orthogonal frequency-division multiplexing \\ \hline
OTFS & orthogonal time-frequency modulation  \\ \hline
PAPR & peak-to-average power ratio \\ \hline
PN & phase noise \\ \hline
QAM & quadrature amplitude modulation \\ \hline
RF & radio-frequency \\ \hline
RMS & root mean square \\ \hline 
SA & subarray \\ \hline
SC & single-carrier \\ \hline
SM & spatial modulation \\ \hline
SNR & signal-to-noise ratio \\ \hline
SPP & surface plasmon polariton \\ \hline
Tbps & terabit-per-second \\ \hline
THz & terahertz \\ \hline
THz-TDS & terahertz time-domain spectroscopy \\ \hline
UE & user equipment \\ \hline
UM-MIMO & ultra-massive MIMO \\ \hline
VLC & visible light communications \\ \hline
WDM & wavelength division multiplexing \\ \hline

 \hline

 \end{tabular}
\end{table}

\section{System Model}
\label{sec:sysmodel}

It is challenging to define a generic system model for THz communications at this early stage. Nevertheless, the use of AoSAs of AEs is most likely to be the norm in future THz systems, as dynamic array gains are crucial for combating the distance problem \cite{Akyildiz8387211}. A typical THz communications system model is illustrated in Fig. \ref{Fig_RF}, where adaptive AoSAs are configured at the transmitting and receiving sides. After the digital-to-analog converter (DAC) and before the ADC, each SA is fed with a dedicated RF chain. Due to high directivity, each SA is detached from its neighboring SAs in a multi-user setting. The role of baseband precoding reduces to defining the utilization of SAs or simply turning SAs on and off. In a point-to-point setup, however, SA paths can be highly correlated due to low spatial resolution. 


We adopt the three-dimensional (3D) UM-MIMO model of \cite{Han8417893,Lin7786122,Sarieddeen8765243,simonTeraMIMO}. The AoSAs consist of $M_t\!\times\!N_t$ and $M_r\!\times\!N_r$ SAs, at the transmitter and the receiver, respectively. Each SA is composed of $Q\times Q$ AEs. Therefore, the overall configuration can be represented as a ``large'' $M_tN_tQ^2\times M_rN_rQ^2$ MIMO system \cite{Sarieddeen8186206}. Such large and near-symmetric doubly-massive MIMO systems \cite{Buzzi8277180} differ from conventional massive MIMO systems. In the latter, large antenna arrays are typically configured at a transmitting base station to serve multiple single-antenna users at the receiver. The distances separating two SAs or two AEs are critical design parameters in reconfigurable settings, as discussed in subsequent sections. We denote these distances by $\Delta$ and $\delta$, respectively. 

THz signal propagation is highly directional (quasi-optical) for three main reasons. First, limited THz reflected components and negligible scattered and diffracted components result in channels dominated by the line-of-sight (LoS) path and assisted by possibly very few NLoS reflected multipath components. Second, high-gain directional antennas are typically used to combat the distance problem instead of omnidirectional antennas {with $\unit[0]{dBi}$ gains}, which further reduces the surviving paths to a single path. Third, the high array gains of beamforming guarantee directional ``pencil beams,'' where typically each SA generates a single beam. Consequently, we assume for the generic system model an LoS transmission over an SC frequency-flat fading channel. The corresponding baseband system model is \vspace{2mm}
\begin{equation}\label{eq:sysmodel}\vspace{2mm}
	\mbf{y} = \mbf{W}^{\Hr}_r \mbf{H}\mbf{W}^{\Hr}_t \mbf{x} + \mbf{W}^{\Hr}_r \mbf{n},
\end{equation} 
where $\mbf{x}\!=\![x_{1}x_{2}\cdots x_{N_s}^{}]^{T}\!\in\!\mathcal{X}^{N_s\times1}$ is the information-bearing symbol vector of components belonging to a QAM constellation $\mathcal{X}$; for example, $\mbf{y}\!\in\!\mathbb{C}^{N_s\times1}$ is the received symbol vector,  $\mbf{H}\!=\![\mbf{h}_{1}\mbf{h}_{2}\cdots \mbf{h}_{M_tN_t^{}}]\!\in\!\mathbb{C}^{M_rN_r\times M_tN_t}$ is the channel matrix, $\mbf{W_t}\!\in\!\mathcal{R}^{N_s \times M_tN_t}$ and $\mbf{W_r}\!\in\!\mathcal{R}^{M_rN_r\times N_s}$ are the baseband precoder and combiner matrices, and $\mbf{n}\!\in\!\mathbb{C}^{M_rN_r\times1}$ is the additive white Gaussian noise (AWGN) vector of power $\sigma^2$.


An element of $\mbf{H}$, $h_{m_rn_r,m_tn_t}$--the frequency response between the $(m_t,n_t)$ and $(m_r,n_r)$ SAs--is thus defined as \vspace{2mm}
\begin{equation}\label{eq:channel}\vspace{2mm}
	h_{m_rn_r,m_tn_t} = \mbf{a}^{\Hr}_r(\phi_r,\theta_r)G_r \alpha_{m_rn_r,m_tn_t} G_t \mbf{a}_t(\phi_t,\theta_t),
\end{equation} 
for $m_r\!=\!1,\cdots,M_r$, $n_r\!=\!1,\cdots,N_r$, $m_t\!=\!1,\cdots,M_t$, and $n_t\!=\!1,\cdots,N_t$, where $\alpha$ is the path gain, $\mbf{a}_t$ and $\mbf{a}_r$ are the transmit and receive SA steering vectors, $G_t$ and $G_r$ are the transmit and receive antenna gains, and $\phi_t$,$\theta_t$ and $\phi_r$,$\theta_r$ are the transmit and receive angles of departure (AoD) and arrival (AoA), respectively ($\phi$'s are the azimuth angles and $\theta$'s the elevation angles). 

{The steering vectors} can be expressed as a function of the transmit and receive mutual coupling matrices, $\mbf{C}_t,\mbf{C}_r\!\in\!\mathcal{R}^{Q^2\times Q^2}$ as $\mbf{a}_t(\phi_t,\theta_t)\!=\!\mbf{C}_t\mbf{a}_0(\phi_t,\theta_t)$ and $\mbf{a}_r(\phi_r,\theta_r)\!=\!\mbf{C}_r\mbf{a}_0(\phi_r,\theta_r)$. By setting $\mbf{C}_t\!=\!\mbf{C}_r\!=\!\mbf{I}_{Q^2}$ ($\mbf{I}_N$ is an identity matrix of size $N$), the effect of mutual coupling is neglected. Such an assumption is especially valid in the plasmonic case when $\delta\!\geq\!\lambda_{\spp}$ \cite{Zakrajsek7928818}, where the SPP wavelength, $\lambda_{\spp}$, is much smaller than the free-space wavelength, $\lambda$. The ideal SA steering vector at the transmitter side can thus be expressed as \cite{Han8417893}
\begin{equation}\label{eq:steering}
	\mbf{a}_0(\phi_t,\theta_t) \!=\! \frac{1}{Q} [e^{j\Phi_{1,1}},\!\cdots\!,e^{j\Phi_{1,Q}},e^{j\Phi_{2,1}},\!\cdots\!,e^{j\Phi_{p,q}},\!\cdots\!,e^{j\Phi_{Q,Q}}]^T,
\end{equation} 
where $\Phi_{p,q}$ is the phase shift that corresponds to AE $(p,q)$, and is defined as 
\begin{multline}\label{eq:shifts}
	\Phi_{p,q}(\phi_t,\theta_t) = \psi_x^{(p,q)}\frac{2\pi}{\lambda}\cos \phi_t \sin \theta_t \\ + \psi_y^{(p,q)}\frac{2\pi}{\lambda}\sin \phi_t \sin \theta_t  + \psi_z^{(p,q)}\frac{2\pi}{\lambda}\cos \theta_t,
\end{multline} 
with $\psi_x^{(p,q)}$, $\psi_y^{(p,q)}$, and $\psi_z^{(p,q)}$ being the coordinate positions of AEs in the 3D space. Furthermore, for analog beamforming per SA, given the target AoDs, $\hat{\phi}_{t},\hat{\theta}_{t}$, the beamforming vector $\hat{\mbf{a}}_t(\hat{\phi}_{t},\hat{\theta}_{t})$ can be similarly defined using
\begin{multline}\label{eq:beamformingshifts}
	\hat{\Phi}_{p,q}(\hat{\phi}_t,\hat{\theta}_t) = \psi_x^{(p,q)}\frac{2\pi}{\lambda}\cos \hat{\phi}_t \sin \hat{\theta}_t \\ + \psi_y^{(p,q)}\frac{2\pi}{\lambda}\sin \hat{\phi}_t \sin \hat{\theta}_t  + \psi_z^{(p,q)}\frac{2\pi}{\lambda}\cos \hat{\theta}_t.
\end{multline} 
The equivalent array response could then be represented as \cite{simonTeraMIMO}
\begin{equation}
    \mbf{a}_{\mathrm{eq}}^{(t)} = \frac{1}{\sqrt{M_tN_t}}\sum^{M_t}_{m_t=1}\sum^{N_t}_{n_t=1}e^{j\frac{2\pi}{\lambda}\left(\Phi_{p,q}(\phi_t,\theta_t)-\hat{\Phi}_{p,q}(\hat{\phi}_t,\hat{\theta}_t)\right)}.
    \label{eq:Aeq_3darray}
\end{equation}
At the receiver side, $\mbf{a}(\phi_r,\theta_r)$, beamforming vector $\hat{\mbf{a}}(\hat{\phi}_{r},\hat{\theta}_{r})$ ($\hat{\phi}_{r},\hat{\theta}_{r}$ are the target AoA), and $\mbf{a}_\mathrm{eq}^{(r)}$ can be similarly defined.


\section{THz-Band Channel Modeling}
\label{sec:channel_mod}

Channel modeling is essential for efficient signal processing in the THz band. Accurate THz channel models should consider the effect of both the spreading loss and the molecular absorption loss and should account for the LoS, NLoS, reflected, scattered, and diffracted signals. Channel modeling approaches are primarily deterministic or statistical \cite{Han8387210}. Although deterministic channel modeling uses computationally extensive ray-tracing techniques to capture site geometry, matrix-based statistical modeling represents each independent sub-channel using a random variable of a specific distribution. Hybrid channel modeling schemes combine the advantages of both approaches, where dominant paths are captured deterministically and other paths are generated statistically \cite{chen2021channel}.

\subsection{Ray-Based THz Channel Modeling}
\label{sec:RT}



Several extensive ray-tracing-based THz propagation measurements have been recently reported. For instance, a unified multi-ray THz-band channel model is proposed in \cite{Han6998944}, which covers the LoS, scattered, reflected, and diffracted paths and is experimentally validated over $\unit[0.06 - 1]{THz}$. Ray-tracing techniques are similarly used to post-process sub-THz channel measurements in \cite{Yu9145314}. In \cite{Moldovan7063462}, a deterministic channel model over $\unit[0.1 - 1]{THz}$ is proposed for LoS and NLoS scenarios, using the Kirchhoff scattering theory and ray tracing. Similarly, ray-based sub-THz channel characterization at $\unit[90 - 200]{GHz}$ is detailed in \cite{gougeon2019sub} using deterministic simulations in indoor office and outdoor in-street scenarios. 


Other THz ray-tracing channel modeling attempts are tailored for the peculiarities of specific use cases. For instance, a ray-tracing channel model at $\unit[300]{GHz}$ is presented in \cite{He7928447}, for close-proximity THz communications, such as for Kiosk downloading. Also, at $\unit[300]{GHz}$, a ray-tracing simulator with calibrated electromagnetic parameters is used in \cite{Yi8903275} for vehicle-to-infrastructure THz communications. For reducing the complexity of ray tracing in UM-MIMO systems, select few virtual paths can be captured between virtual transmitting and receiving points, the response of which gets mapped to actual pairs of transmitting and receiving AEs.

\subsection{Statistical THz Channel Modeling}
\label{sec:statistical}

Several statistical THz channel modeling alternatives to time-consuming and complex ray-tracing models in fixed geometries have been attempted. For example, by developing a wideband channel sounder system at $\unit[140]{GHz}$, indoor wideband propagation and penetration measurements for common building materials are reported in \cite{xing2018propagation}. In \cite{xing2019indoor}, indoor measurements and models for reflection, scattering, transmission, and large-scale path loss are provided by the same group for mmWave and sub-THz frequencies; a 3D statistical indoor channel model is further reported in \cite{ju20203}. A lower reflection loss is noted at higher frequencies in indoor drywall scenarios (stronger reflections). In contrast, the partition loss increases due to more prominent depolarizing effects. Furthermore, recent measurement-based models are reported by the same group for sub-THz urban microcell \cite{xing2021propagation}, indoor office \cite{xing2021millimeter}, and \cite{xing2021terahertz} space scenarios. {Natural isolation is noted} between terrestrial networks and surrogate satellite systems and between terrestrial mobile users and co-channel fixed backhaul links. 

The statistical characterization of three channel bands between $\unit[300]{THz}$ and $\unit[400]{THz}$ is presented in \cite{khalid2019statistical} based on a broad set of measurements in LoS and NLoS environments and including spatial and temporal variations. The large-scale losses are modeled using the single slope path loss model with shadowing, where variations due to shadowing are normally distributed. Metal, wood, and acoustic ceiling panels are confirmed as good reflectors. Strong multipath components reduce the coherence bandwidth in indoor environments significantly, and high channel correlation is retained. With a virtual antenna array technique, the same testbed is exploited in \cite{Khalid7562539} to demonstrate $2\!\times\!2$ THz LoS MIMO channels. In \cite{nguyen2020large}, channel measurements agree closely with the new radios (NR) channel model of the Third Generation Partnership Project (3GPP) for a specific indoor scenario.

Another stochastic indoor $\unit[300]{GHz}$ spatio-temporal channel model is introduced in \cite{Priebe6574880} that considers parameters such as polarization, ray amplitudes, times of arrivals, angles of arrivals and departures, and path-specific frequency dispersion. Furthermore, THz channel modeling via a mixture of gamma distributions is proposed in \cite{Tekbiyik9368251}. In other notable studies, LoS broadband THz channel measurements are reported in \cite{Mirhosseini8761281} when using convex lenses at the transmitter, coupled with collimating lenses at the receiver. A beam domain channel model is also introduced in \cite{You7913686}. With large numbers of base stations and users, and given that several wavelengths typically separate users at high frequencies, the beam-domain channel elements are statistically uncorrelated \cite{6542746Adhikary}, and their envelopes are independent of frequency and time. Moreover, a geometric-based stochastic time-varying model at $\unit[110]{GHz}$ is proposed in \cite{Chen9013865} for THz vehicle-to-infrastructure communications. 


{
A comprehensive statistical simulator of 3D end-to-end wideband UM-MIMO THz channels has recently been introduced in \cite{simonTeraMIMO}. The simulator, called TeraMIMO, models THz channel statistics such as coherence time, coherence bandwidth, Doppler spread, and root mean square (RMS) delay spread. TeraMIMO generates frequency-selective and time-variant THz channels for different communication distances, ranging from nanocommunications to short-range indoor and outdoor communications and LoS links of hundred of meters. TeraMIMO also accounts for LoS-dominant and NLoS-assisted scenarios. By adopting the same AoSA architecture of our system model, TeraMIMO studies the resultant spatial effects of misalignment, spherical wave propagation, and beam split in wideband THz channels. Such tools can catalyze THz research, especially if they continue to follow the latest results from measurement-based THz channel modeling campaigns.
}

\subsection{Effect of Scattering}
\label{sec:scattering}

The electromagnetic roughness of surfaces increases at higher frequencies, causing diffuse scattering and increased backscattering (at lower incident angles) \cite{Bystrov8904694}. The effect of scattering on the reflection coefficient in THz communications is studied in \cite{Ju8761205}. The scattered power increases with frequency and surface roughness relative to the reflected power, where smooth surfaces (like drywall) can be modeled as reflective surfaces. In \cite{Sheikh8859609}, diffuse scattering in THz massive MIMO channels is studied by developing a hybrid modeling approach for 3D ray-tracing simulations by assuming realistic indoor environments over the $\unit[0.3 - 0.35]{THz}$ band. The channel capacity of indoor massive MIMO channels is calculated by assuming different surface roughnesses for LoS and NLoS scenarios. Scattering can be leveraged to achieve a trade-off between rich multipath and high received power in THz massive MIMO. Scattering can thus enhance the spatial multiplexing gains. Diffuse scattering can also be leveraged to identify surface types. 

In \cite{amarasinghe2019scattering,renaud2019terahertz}, it is argued that THz beams are more susceptible to snow than rain, suffering higher losses under an identical fall rate Mie theory approach for electromagnetic radiation. THz-band rain-induced co-channel interference is studied in \cite{Juttula8756966}, using the bistatic radar and the Mie scattering theory, and assuming first-order multiple scattering. The overall interference levels from rain are significantly lower in the THz band ($\unit[20]{dB}$ difference between $\unit[300]{GHz}$ and $\unit[60]{GHz}$), except when the receiver is very close to the LoS (forward-oriented scattering at high frequencies). Furthermore, in \cite{xing2021terahertz}, rain attenuation measurements at $\unit[140]{GHz}$ reveal that communications above $\unit[70]{GHz}$ are not further impacted by rain.


\subsection{Effect of Blockage}

Statistical modeling can also be used to study the effect of blockage, which is significant at higher frequencies. For instance, NLoS THz channel modeling is conducted in a generic stochastic approach in \cite{hossain2017stochastic} by assuming rectangular geometry and accounting for variable densities of reflecting objects (single reflection components) and blocking obstacles. THz signals are more sensitive to blockages than mmWave signals. Human blockage in indoor THz communications is studied in \cite{Bilgin8757158}, where adding extra antennas to account for blocked streams is considered. The average network throughput and coverage probability in the presence of indoor blockage effects resulting from human bodies and walls are further studied in \cite{Wu9247469}. 

The dynamic blockages caused by moving humans are incorporated into the 3D THz channel model in \cite{shafie2020coverage}, where deploying antennas at high altitudes would help prevent blockages. Furthermore, the effect and mitigation of blockage in THz relay systems are addressed in \cite{stratidakis2020relay,Boulogeorgos9217121}. Moreover, blockages can arise at the transmitter due to the condensation of particles, and such blockages can be mitigated using compressive sensing techniques \cite{Eltayeb7841677,Eltayeb8248776}. A similar remote array diagnosis technique that mitigates antenna blockage without the need for full channel state information (CSI) is reported in \cite{medina2020millimeter}.




\begin{figure*}[ht]
\begin{equation}
\begin{aligned}
\label{abs_coeff}
    &\! \mathcal{K}(f) \!=\! \sum_{i,g\!}\! \frac{p}{p_0}\!\frac{T_{STP}}{T}\! \frac{p}{RT} q^{i,g\!}\! N_A S^{i,g\!}\! \frac{f}{f_{c0}^{i,g\!}\!+\!\delta^{i,g\!}\frac{p}{p0}} \frac{\text{tanh}\left(hcf\!/2K_BT\right)}{\text{tanh}\!\left(hc\left(\!f_{c0}^{i,g\!}\!+\!\delta^{i,g\!}\!\frac{p}{p0}\!\right)\!/2K_BT\right)} \frac{\left[(1\!-\!q^{i,g\!})\alpha_0^{\mathrm{air}}\!+\!q^{i,g\!}\alpha_0^{i,g\!}\right]\left(p/p_0\right)\left(T_0/T\right)^\gamma}{\pi}\frac{f}{f_{c0}^{i,g\!}\!+\!\delta^{i,g\!}\frac{p}{p0}} \\ & \left(\!\frac{1}{\left(f\!-\!\left(f_{c0}^{i,g\!}+\delta^{i,g\!}\frac{p}{p0}\right)\right)^2\!+\!\left(\left[\left(1\!-\!q^{i,g\!}\right)\alpha_0^{\mathrm{air}}\!+\!q^{i,g\!}\alpha_0^{i,g\!}\right]\!\left(\frac{p}{p_0}\right)\!\left(\frac{T_0}{T}\right)^\gamma\right)^2} \!+\! \frac{1}{\left(f\!+\!\left(f_{c0}^{i,g\!}+\delta^{i,g\!}\frac{p}{p0}\right)\right)^2\!+\!\left(\left[\left(1\!-\!q^{i,g\!}\right)\alpha_0^{\mathrm{air}}\!+\!q^{i,g\!}\alpha_0^{i,g\!}\right]\!\left(\frac{p}{p_0}\right)\!\left(\frac{T_0}{T}\right)^\gamma\right)^2}\!\right)
\end{aligned}
\end{equation}
\end{figure*}

\subsection{The Molecular Absorption Effect}
\label{sec:abs}

The path loss of a THz signal in the presence of water vapor is dominated by spikes that represent molecular absorption losses originating at specific resonant frequencies due to excited molecule vibrations. Higher densities of absorbing molecules strengthen and widen the peaks (broadening of absorption lines). The spectrum is divided into smaller windows (sub-bands) of tens or hundreds of GHz because of these lines. The windows are distance-dependent because some spikes only become significant at specific distances (by increasing the distance from 1 to 10 meters, the transmission windows are reduced by order of magnitude \cite{Hossain8761547}). Hence, variations in the communication distance affect both the available bandwidths and the path loss, where the available bandwidth shrinks at higher frequencies. 


The LoS path gain as a function of absorption is expressed as
\begin{align}\label{eq:LoS}
	\alpha_{m_rn_r,m_tn_t}^{\LoS} =  & \frac{c_0}{4\pi f d_{m_rn_r,m_tn_t}} \\ & \times e^{ -\frac{1}{2} \mathcal{K}(f) d_{m_rn_r,m_tn_t} }  e^{  -j \frac{2\pi f}{c} d_{m_rn_r,m_tn_t}},
\end{align} 
where $d_{m_rn_r,m_tn_t}$ is the distance between the transmitting and receiving SAs, $\mathcal{K}(f)$ is the absorption coefficient, $f$ is the frequency of operation, and $c_0$ is the speed of light in vacuum. $\mathcal{K}(f)$ is derived in \cite{Jornet5995306} as a summation over contributions from isotopes ($i\!\in\! {1,\cdots,I}$) of gases ($g\!\in\! {1,\cdots,G}$) that constitute a medium. The construction in \cite{Jornet5995306} uses radiative transfer theory for insight into the physical meaning of the corresponding equations as a function of temperature, system pressure, and absorption cross-section.

{We compile the overall} equation for $\mathcal{K}(f)$ in \eqref{abs_coeff}, where  $T$ is the system temperature, $T_0$ is reference temperature, $K_B$ is the Boltzman constant, $h$ is the Planck constant, $R$ is the gas constant, $N_A$ is the Avogadro constant, $p$ is the system pressure, $p_0$ is the reference pressure, $q^{i,g}$ is the mixing ratio of gas $(i,g)$, $f_{c0}^{i,g}$ is the resonant frequency at the reference pressure, $\gamma$ is the temperature broadening coefficient, $\delta^{i,g}$ is the linear pressure shift of gas $(i,g)$, $S^{i,g}$ is the line intensity, $\alpha^{\mathrm{air}}_0$ is the broadening coefficient of air, and $\alpha^{i,g}_0$ is the broadening coefficient of gas $(i,g)$. All these parameters can be extracted from the high-resolution transmission molecular absorption database (HITRAN) \cite{gordon2017hitran2016}. However, this model is complex and challenging to track analytically.

Because water vapor dominates the absorption losses at high frequencies, simplified yet sufficiently accurate models for molecular absorption loss are developed in \cite{Kokkoniemi8568124,kokkoniemi2020line} and used in \cite{Boulogeorgos8445864,Papasotiriou8580934,Boulogeorgos8417891}. These models are built using a database approach by fitting the absorption line shape functions to the actual responses. In the first model, tailored for the $\unit[0.275 - 0.4]{THz}$ band, the absorption coefficient is approximated as
\begin{equation}
\mathcal{K}(f)=K_{1}(f, v)+K_{2}(f, v)+K_3(f),
\end{equation}
where
\begin{align}
K_{1}(f, v) & = \frac{A(v)}{B(v)+\left(\frac{f}{100 c}-c_{1}\right)^{2}}, \\
K_{2}(f, v) & = \frac{C(v)}{D(v)+\left(\frac{f}{100 c}-c_{2}\right)^{2}}, \\
K_3(f) & = \rho_{1} f^{3}+\rho_{2} f^{2}+\rho_{3} f+\rho_{4},\vspace{2mm}
\end{align}
with ${v}$ being the volume mixing ratio of water vapor. The rest of the coefficients and functions are detailed in \cite{Kokkoniemi8568124}. The updated model in \cite{kokkoniemi2020line} is valid over an extended sub-THz range ($\unit[100-450]{GHz}$), accounting for more absorption spikes. Both approximations are sufficiently accurate for links up to $1$ kilometer (Km) under standard atmospheric conditions. Although such approximations reduce the computational complexity and are easy to track analytically, the exact HITRAN-based absorption model is still favored, especially in very high SNR settings. In the context of joint signal processing for communications and sensing, where exact knowledge of medium components is sought (Sec. \ref{sec:sensing}), precise HITRAN-based molecular absorption modeling is crucial.

\begin{figure}[t]
  \centering
  \includegraphics[width=0.5\textwidth]{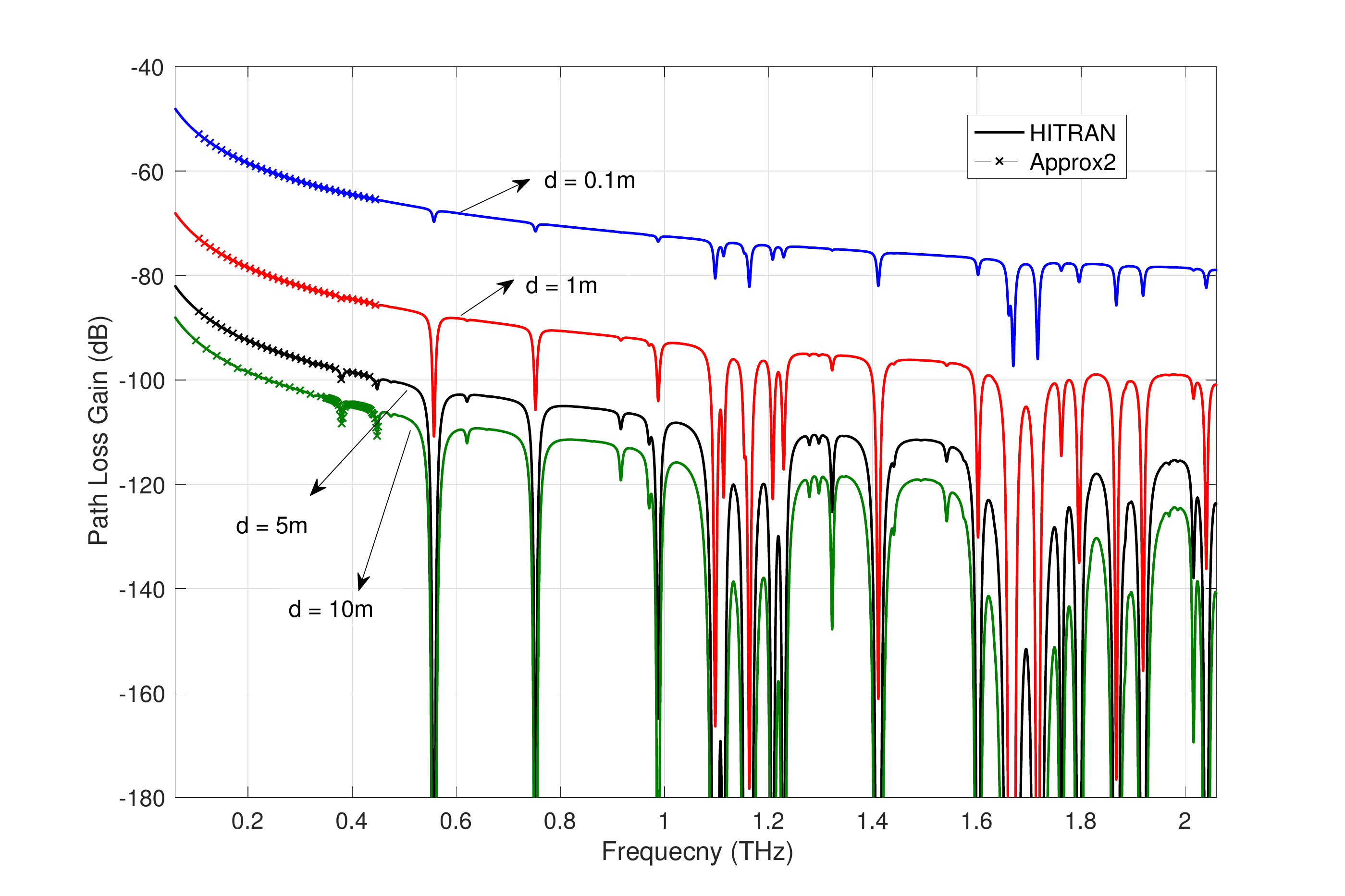}\vspace{2mm}
 \caption{Comparison between path gains with molecular absorption computed via HITRAN (exact) and approximated (100-450 GHz): 298.15 Kelvin, 1 atm, composition (N2 = 76.5450, O2 = 20.946, H2O = 1.57, CO2 = 0.033, CH4 = 0.906), and relative humidity 50\%.}
  \label{fig:path_loss_d}
\end{figure}

The distance-dependent path loss is illustrated in Fig. \ref{fig:path_loss_d}, which plots the total path loss (i.e., the spreading and the molecular losses) as a function of frequency--increasing the communication distance results in more severe losses. Three spectral windows are noted between path loss peaks below $\unit[1]{THz}$; at medium ranges and frequencies higher than $\unit[1]{THz}$, the spectrum becomes more fragmented, where the window widths depend on both the center frequency and the communication distance.

\subsection{Multipath THz Channels}
\label{sec:multipath}


Despite only considering an LoS-dominant scenario in our system model, a multipath channel can arise in several THz communications scenarios, especially indoors, where lower antenna gains can be tolerated. Nevertheless, it is safe to assume that a THz channel is sparser than a mmWave channel. For instance, only five multipath components survive at $\unit[0.3]{THz}$ in a $256\times256$ UM-MIMO system, 32.5\% lower than the number of multipath components in the same system at $\unit[60]{GHz}$ \cite{Yan8756936}. On average, the path gain difference between LoS and NLoS paths is $\unit[15]{dB}$ higher in THz systems compared to that in mmWave systems \cite{Han8417893}. Consequently, the angular spread of indoor THz channels is much smaller than that at lower frequencies \cite{Lin7786122}. Indoor THz multipath components \cite{Priebe6574880,Lin7037398,Lin7116524,simonTeraMIMO} can be modeled using the Saleh-Valenzuela (S-V) channel model \cite{Saleh1146527}, where the channel response within a time margin $\T_s$ is expressed as
\begin{align}
     h^{\NLoS}_{m,n} =& \sum_{v=0}^{N_{\clu}-1} \sum_{u=0}^{N_{\ray}^{(v)}} \mathbf{a}_{r}^{\mathcal{H}}\left(\phi^{(u,v)}_{r}, \theta^{(u,v)}_{r}\right) G_{r}\left(\phi^{(u,v)}_{r}, \theta^{(u,v)}_{r}\right) \\ 
    & \alpha^{\NLoS (u,v)}_{m,n} G_{t}\left(\phi^{(u,v)}_{t}, \theta^{(u,v)}_{t}\right) 
     \mathbf{a}_{t}\left(\phi^{(u,v)}_{t}, \theta^{(u,v)}_{t}\right),
\end{align}
where $N_{\clu}$ is the number of path clusters and $N_{\ray}^{(v)}$ is the number of paths within the $\nth{v}$ cluster. Each path can have random angles of departure and arrival within a beam region. The expectation of path gain magnitude is
\begin{equation}
\mathsf{E}\left[\left|\alpha^{\NLoS (u,v)}_{m,n}\right|^{2}\right]= \left(\frac{c_0}{4 \pi f d_{m,n}}\right)^{2}\ e^{-\mathcal{K}(f) d_{m,n}} e^{-\frac{\tau_{v}}{\Gamma}} e^{-\frac{\bar{\tau}_{v,u}}{\gamma}},
\end{equation}
where $\tau_{v}$ and $\bar{\tau}_{v,u}$ are the times of arrival that follow an exponential or paraboloid distribution and ${\Gamma}$ and ${\gamma}$ are the clusters and ray decay factors, respectively. The angles of departure and arrival are calculated as
\begin{equation}\begin{array}{ll}
\phi^{(u,v)}_{t}=\bar{\varphi}_{t}^{(u)}+\varphi_{t}^{(u,v)}, & \phi^{(u,v)}_{r}=\bar{\varphi}_{r}^{(u)}+\varphi_{r}^{(u,v)}, \\
\theta^{(u,v)}_{t}=\bar{\vartheta}_{t}^{(u)}+\vartheta_{t}^{(u,v)}, & \theta^{(u,v)}_{r}=\bar{\vartheta}_{r}^{(u)}+\vartheta_{r}^{(u,v)},
\end{array}\end{equation}
where $\bar{\varphi}_{t}^{(u)}$/$\bar{\varphi}_{r}^{(u)}$ and $\bar{\vartheta}_{t}^{(u)}$/$\bar{\vartheta}_{r}^{(u)}$ are the uniformly distributed cluster azimuth and elevation angles of departure/arrival, and  $\varphi_{t}^{(u,v)}/\varphi_{r}^{(u,v)}$ and $\vartheta_{t}^{(u,v)}/\vartheta_{r}^{(u,v)}$ are the ray azimuth and elevation angles of departure/arrival. The latter can follow a zero-mean second-order Gaussian mixture model \cite{simonTeraMIMO}.

\subsection{Wideband Channels and Beam Split}
\label{sec:beamsplit_effect}

Frequency selectivity due to multipath or molecular absorption is more significant in wideband THz channels supported in several scenarios. Ultra-wideband pulse-based modulations are serious candidates for THz communications in which short pulses in time span the entire THz range in frequency \cite{Tsujimura8877200,Tsujimura8659776}. In such wideband scenarios, a deterministic frequency-selective fading, not captured by Rayleigh and Ricean models, arises due to molecular absorption. This fading results in delayed signal components in a wideband multipath scenario. {Hence, group velocity dispersion} (GVD) is another issue that arises in impulse radio THz communications due to frequency-dependent refractivity in the atmosphere. GVD becomes limiting at specific link distances, atmospheric water vapor densities, and channel bandwidths \cite{mandehgar2014experimental}. This phenomenon also results in inter-symbol interference (ISI) as data bits spread out of their assigned slots and interfere with neighboring slots. Therefore, dilating bit slots can solve this problem at the expense of the data rate. In \cite{strecker2019compensating}, the atmospheric GVD of THz pulses ($\unit[0.2 - 0.3]{THz}$) is compensated for using stratified media reflectors.

Beam split is another critical phenomenon that arises in ultra-broadband THz signals with large fractional bandwidths (ratio between bandwidth and central frequency). The beam splits when different THz path components at different subcarriers squint into different spatial directions, resulting in array gain loss~\cite{dai2021delay}. Beam squint can occur due to frequency-independent delays in analog beamforming phase shifters when the same phase shift is applied to different frequencies. The difference between $f_k$, the frequency of a subcarrier in a multi-carrier system, and $f_c$, the center frequency, is significant at THz frequencies. Therefore, the paths split into different spatial directions within an ultra-broadband bandwidth. The angle domain beam split is expressed as \cite{simonTeraMIMO}
\begin{equation}
    \label{eq:spat_dir_subc_fc}
    \Phi_{p,q}(\phi_t,\theta_t,f_k) = \frac{f_k}{f_c} \Phi_{p,q}(\phi_t,\theta_t,f_c) .
\end{equation}

The large UM-MIMO AoSAs also add a beam split effect when the signal propagation time between SAs introduces a frequency-dependent phase shift that biases the estimated angle of arrival in steering vectors \cite{dai2021delay}. The highly narrow beamwidths under UM-MIMO beamforming worsen the beam split effect. The receiving antenna array might often be larger than the received beamwidth, which indicates that a path is not visible to all the antennas in the array. This phenomenon causes non-stationarity \cite{de2019non} and variations in time of arrival, angle of arrival (and angle of departure), and received amplitudes across the antenna array. The variation in the time of arrival causes ISI (not only a phase shift) in wideband scenarios \cite{Wang8354789}. 

Beam split can be mitigated in the digital domain of hybrid architectures \cite{han2021hybrid,Zhang9130760}. In \cite{dovelos2021channel}, a beam squint mitigation scheme based on introducing true-time-delays over uniform planar arrays is investigated for channel estimation and hybrid combining. In \cite{Perera8974228}, the use of radix-2 self-recursive sparse factorizations of delay Vandermonde matrices as alternatives to fast Fourier transform is robust to beam split effects. Furthermore, wideband beam zooming schemes based on beam tracking are proposed to combat beam split, such as the use of delay-phase precoding \cite{Tan9348222,Chen9158439,Tan9348222}.

\subsection{Spherical Wave Model}
\label{sec:swm_effect}

The near- and far-field considerations are also critical at THz frequencies. The plane wave propagation model applies to far-field scenarios where the distance between the transmitter and the receiver is greater than or equal to the Rayleigh distance of the antenna array \cite{bohagen2009spherical}, or when the distance is comparable to the array dimensions \cite{Bjornson9184098}. At lower microwave and mmWave frequencies, this distance is less than $\unit[0.5]{m}$ and $\unit[5]{m}$ for an array size of $\unit[0.1]{m}$ and an operating frequency of $\unit[6]{GHz}$ and $\unit[60]{GHz}$, respectively. However, this distance grows to approximately $\unit[40]{m}$ at $\unit[0.6]{THz}$, which is greater than most achievable THz communication distances, confirming the importance of the spherical wave propagation models. The spherical wave model should be considered when the distance is within the Fresnel region \cite{balanis2016antenna}:
\begin{equation}
     0.62\sqrt{\frac{D^3}{\lambda}} \le d < \frac{2D^2}{\lambda},
\end{equation}
where $D$ is the overall maximum antenna dimension (a slightly different MIMO near-field definition is adopted in \cite{Bjornson9184098}). Spherical waves can be modeled at AE or SA levels. In compact device technologies, such as plasmonics, an SA-level spherical wave model can be sufficient for most communication distances. Retaining the plane wave assumption at the AE level has the advantage of generating a simple and convenient equivalent array response. The spherical signal wave considerations are at the distance and angle levels--the distance between two SAs in the far-field should be calculated by accounting for the curvature, and the AoD/AoA differ for different SAs.

\subsection{Antenna Gains}
\label{sec:ant_g}

Directional antennas are essential for overcoming the high THz propagation losses. A simplified ideal sector model for antenna gains, $G_t$ at the transmitter and $G_r$ at the receiver, can be expressed as \cite{Lin7036065}:
\begin{equation}
G=\left\{
    \begin{array}{c l}\sqrt{G_0},& \forall {\phi}\in[\phi_{\mathrm{min}},\phi_{\mathrm{max}}],\forall{\theta}\in[\theta_{\mathrm{min}},\theta_{\mathrm{max}}],\\0,&\mathrm{otherwise}.\end{array}
\right.
 \label{eq:antenna_gain_sector_model}
\end{equation}
This model can be applied for both LoS and NLoS components with the corresponding azimuth and elevation angles. For highly directional antennas, $G_0$ can be approximated as~\cite{balanis2016antenna,xia2019link}
\begin{equation}
    G_0 = \frac{4\pi}{\psi_A} \approx \frac{4\pi}{\psi_{\phi}\psi_{\theta}},
    \label{eq:antenna_gain_approx}
\end{equation}
where $\psi_A$ is the beam solid angle, and $\psi_{\phi}$ and $\psi_{\theta}$ are the half-power beamwidths (HPBWs) in the azimuth and elevation planes, respectively. The antenna sectors are small under antenna directivity. For example, HPBW azimuth/elevation-plane angles $\psi_\phi\!=\!\psi_\theta\!=\!27.7^{\circ}$ only result in a $\unit[17.3]{dBi}$ gain \cite{xia2019link}. However, much higher gains are required in medium-distance THz communications. Array gains can complement antenna gains.  For instance, a massive number of THz-operating antennas can be fit into a few square millimeters. 


\subsection{Effect of Misalignment and Impairments}
\label{sec:misalignment}

The performance of THz communications systems severely deteriorates under the effect of misalignment and hardware impairments. Misalignment occurs when the transmitter and receiver do not precisely point to each other, a highly probable scenario with narrow THz beams~\cite{Cacciapuoti8387213}. The joint effect of misalignment and hardware impairments on THz communications is studied in \cite{Boulogeorgos8610080}, including in-phase and quadrature imbalance (IQI) and non-linearities. The study models all impairments as Gaussian noise components and accounts for operation and design parameters alongside environmental parameters; it also introduces a misalignment fading model. However, the adopted model is based on the optical receiver's intensity fluctuation derived in~\cite{farid2007outage}. 

{Alternative THz- and} UM-MIMO-specific misalignment models need to be derived, by complementing the optical model with approximations of the effective radius of the receiving AoSA area, such as in \cite{simonTeraMIMO}. The research in \cite{Boulogeorgos8610080} is extended in \cite{Boulogeorgos8932597} to capture the error analysis of mixed THz-RF wireless systems. The use of high-directivity antennas in THz systems results in small transceiver antenna beamwidths, which provide higher antenna gains but cause pointing errors and loss of connection. Due to the symmetry of the beam, the misalignment fading component depends only on the radial distance \cite{kokkoniemi2020impact}. Earlier attempts to capture this misalignment effect are reported in \cite{Han7579223,Priebe6205849,Ekti8088634}. The effect of small-scale mobility on THz systems is studied in \cite{petrov2018effect,petrov2020capacity}. Simple shakes or rotations due to user equipment (UE) mobility can result in beam misalignment and SNR degradation, which result in a loss in communication time due to extra beam-search mechanisms. Therefore, there exists a trade-off between antenna directivity and capacity.

Misalignment can be modeled by expressing the effective channel coefficient between two SAs in terms of three components \cite{Boulogeorgos8610080} as
$h_\mathrm{eff} \!=\! h  h_{\mathrm{ma}} h_{\mathrm{st}}$, where $h$ is expressed in \eqref{eq:channel}, $h_{\mathrm{ma}}$ is the misalignment fading, and $h_{\mathrm{st}}$ is the stochastic path gain (can be neglected or modeled as an $\alpha-\mu$ process depending on the scenario). Due to beam symmetry, misalignment fading depends primarily on the pointing error, which can be expressed in the form of a radial distance $r$ between the transmission and reception beams at a communication distance $d$:
\begin{equation}
h_{\mathrm{ma}}(r ; d) \approx A_0 \exp \left(-\frac{2 r^{2}}{w_{\mathrm{eq}}^{2}}\right),
\end{equation} 
where $A_0$ is the fraction of power collected at the receiver, and $w_{\mathrm{eq}}$ is the equivalent beamwidth.
Hardware imperfections in both the transmitter and the receiver can be modeled as two additional distortion noises. The modified system model under impairments is approximated as \vspace{2mm}
\begin{equation}\label{eq:sysmodel_imp}
	\mbf{y} =  \mbf{H}(\mbf{x}+\mbf{n}_t) + \mbf{n}_f + \mbf{n},
\end{equation}
where $\mbf{n_t}\!\in\!\mathbb{C}^{M_t N_t\times1}$ and $\mbf{n_f}\!\in\!\mathbb{C}^{M_r N_r\times1}$ are two complex Gaussian distortion noise vectors at the transmitter and the receiver, with noise variances $\eta_{t}^{2} \dot{p}$ and $\eta_{r}^{2} \dot{p}|h|^{2}$, respectively, with $\dot{p}$ being the average transmitted power and $\eta_{t}$ and $\eta_{f}$ being the impairment coefficients \cite{Boulogeorgos8610080}.





\subsection{THz Noise Modeling}
\label{sec:noise_mod}

Accurate noise models are essential for understanding the behavior of THz systems. Although stochastic models for the electronic noise at THz receivers are still lacking, in \cite{Sen8815595}, two primary sources of noise are noted: (1) thermal noise, which arises at the receiver multiplier and mixer chains, and (2) absorption noise that is channel-induced due to water vapor molecules. The corresponding histogram of the measured noise follows a Gaussian distribution, in accordance with the noise behavior at lower frequencies, as opposed to shot noise in optical receivers. In \cite{kokkoniemi2016discussion}, the transmission-induced noise due to molecular absorption is discussed in more detail. The researchers differentiate between multiple models, most of which are based on the antenna temperature generated by the absorbed energy. However, this molecular absorption noise model has not yet been validated by measurements; it uses sky noise as a basis, which might overestimate the level of the self-induced noise. {In \cite{Hoseini8269042,Hoseini9275389}, molecular absorption is} assumed to result in a rich scattering environment because absorptions are followed by re-radiations with minimal frequency shifts. Nevertheless, such coherent re-radiations can be more realistically lumped in a generic absorption noise factor  \cite{Jornet5995306,kokkoniemi2016discussion}. 

The channel-induced component dominates the overall noise in pulse-based systems (especially in low-noise graphene-based electronic devices \cite{pal2009ultralow}), and is colored over frequency. The total noise power at a distance $d$ can be expressed as \vspace{2mm}
\begin{equation}\label{eq:noise1}
	\sigma^2 =  K_B \int_B T_{\noise}(f,d) df,
\end{equation}
where $T_{\noise} = T_{\sys} + T_{\mol} + T_{\mathrm{other}}$, $T_{\sys}$ and $T_{\mol}$ are the system electronic noise temperature and molecular absorption noise, respectively, and \vspace{2mm}
\begin{equation}\label{eq:noise2}
	T_{\mol}(f,d) =  T_0(1-e^{-\mathcal{K}(f)d}).
\end{equation}
In a carrier-based system with perfect frequency planning over absorption-free spectra, the effect of the channel-induced noise can be minimized. This is particularly true at shorter communication distances, where propagation losses dominate. At larger distances and higher frequencies, however, molecular absorption losses can overcome propagation losses. An additional low-frequency noise component exists at the transmission chain and power supply.


In addition to these noise sources, the effect of phase noise (PN) at THz frequencies should be considered \cite{Scheytt9166347}. PN is caused by time-domain instability (jitter), which produces random rapid, short-term fluctuations in the phase. Precise THz-specific PN measurements are still lacking, despite some modeling attempts \cite{tervo20205g,Bicais9013189}. PN can be modeled by the superposition of Wiener and Gaussian noises. 
Furthermore, PN is typically accompanied by strong phase impairment and carrier frequency offset (CFO)--both result from poorly performing high-frequency oscillators (more of a problem in multi-carrier systems \cite{Yuan9139316}). Novel signal processing techniques and optimized modulation/demodulation schemes can overcome these impairments and achieve PN robustness. THz CFO can be estimated via an under-sampling approach with narrow-band filtering and coprime sampling \cite{Song8488301}. Other notable studies that account for PN and impairments at the sub-THz band include \cite{Saad9145359} on MIMO techniques, \cite{Bicais9052917} on SC transceivers, \cite{Neshaastegaran9310007} on OFDM systems, and \cite{Gougeon9135258} on mesh backhaul capabilities.

\section{Analysis Frameworks and Testbeds}
\label{sec:analysis}

The more accurate the THz channel models get, the more insight we have on the achievable gains of THz systems. These gains can be captured via theoretical performance analysis frameworks and can be verified via experimental testbeds. In this section, we summarize recent results on both.

\subsection{Fundamental Performance Analysis Frameworks}
\label{sec:demos}
 
Early THz channel capacity studies are reported in \cite{Jornet5501885} in the context of nanonetworks, where numerical results are generated for different molecular compositions and power allocation schemes, with emphasis on pulse-based modulation. In \cite{Han6998944}, a thorough analysis of THz channel characteristics is presented, where the variability of spectral window widths over communication distances is first observed. Water-filling power allocation can achieve more than 75 Gbps with $\unit[10]{dBm}$ transmit power over $\unit[0.06 - 1]{THz}$. The study further illustrates that in multipath scenarios, the RMS delay spread is distance- and frequency-dependent, and the coherence bandwidth is less than $\unit[5]{GHz}$ (decreases with longer distance and lower carrier frequencies). Furthermore, the spacings between transmissions are affected by the increased temporal broadening effects at higher frequencies, wider pulse bandwidths, and longer distances. Therefore, distance-adaptive techniques are suggested alongside multi-carrier schemes. 

Theoretical performance analysis of conventional THz communications systems is also progressing. For instance, in \cite{Papasotiriou8580934}, the two-path channel characteristics over $\unit[275-400]{GHz}$ are analyzed in terms of SNR and ergodic capacity by considering channel characteristics such as frequency selectivity, path-loss, and atmospheric conditions. The analysis also accounts for transceiver parameters such as antenna gains and transmit power. By assuming the signal and noise to be jointly Gaussian, classic Shannon results for coherent reception are used for capacity estimation.
{This approach is employed} in several other studies: in \cite{Jornet5995306} to compute the capacity of THz nanosensor networks, in \cite{boronin2014capacity} to capture the relation between transmission distance and absorption-free (transparency) windows, and in \cite{Schneider6155633} to study the data rates of fixed THz-links. Such studies are also extended in \cite{Xu6846379} in the context of reconfigurable MIMO systems. Furthermore, the effect of both deterministic (molecular absorption) and random (atmospheric turbulence and pointing errors) factors on the bit-error-rate performance and capacity of LoS THz links is studied in \cite{taherkhani2020performance}, where the log-normal, gamma-gamma, and exponentiated Weibull channel models are used.


THz systems are also susceptible to transceiver imperfections. In \cite{Papasotiriou9039743}, it is argued that in the presence of PN and misalignment, the outage performance is not significantly enhanced by higher transmit power, and lower-order modulations might be required. The effect of local oscillator hardware impairments can also be more severe than misalignment issues. The performance of THz systems is studied under the joint impact of PN and misalignment fading in \cite{papsotiriou2019performance}, and under the joint impact of PN and amplifier non-linearities in \cite{schenk2008rf}. When lumped into PN at the local oscillator, the impact of these errors is examined under different transceiver architectures in \cite{Aminu8645412}. 


The achievable data rates of indoor THz systems are studied in \cite{Moldovan7996402}, where a single frequency network is advocated, and the corresponding ISI due to channel dispersion considered, and the effect of the density of access points on performance is studied. Similarly, in \cite{Yao8849958}, the indoor interference and coverage under beamforming are studied. An analytical model for the distribution of indoor access points at different blocks of the THz spectrum is proposed in \cite{singh2020analytical}. 
{In THz indoor NLoS scenarios}, the total received signal is accumulated predominantly  from diffuse reflections \cite{Sheikh8888704}. Small-scale mobility in indoor THz scenarios is also studied in \cite{singh2019parameter} as a function of several variables, such as frequency windows, beamwidths,  distance, humidity, mobility type, and antenna placement--there exist optimal beamwidths for specific mobility types and AP placement strategies. In \cite{Petrov7820226}, the interference in mmWave and THz systems under blockage and directional antennas is studied.

\subsection{Performance Analysis for Novel Use-Cases}
\label{sec:demosSG}

It is crucial to study the system-level resource allocation considerations of THz communications. The purpose of such studies is to decide on the optimal use of THz signals in a future communications system, whether in the backhaul of networks to alleviate data rate bottlenecks or in the access domain to deliver novel communication and sensing applications. The performance under different medium conditions (snow, rain) should be studied for backhaul links, comparing THz links to other backhaul solutions. In the access domain, a stochastic geometry approach \cite{Kokkoniemi7880713} can be used, as an alternative to time-consuming and error-prone heavy simulations, to capture the outage probability and optimize resource allocation in dense deployments under interference and blockages. 
{Despite THz signal directionality}, inter-cell interference can still arise in dense deployments of THz base stations that overcome the high path and molecular absorption losses. Stochastic geometry can introduce a mathematically compliant formulation for inter-cell interference, which is extremely difficult under other approaches. Stochastic geometry at the THz band should account for a practical blockage model, highlighting the impact of the penetration loss on the communication quality of the THz channel in indoor and outdoor environments. The impact of the high molecular absorption loss, spreading loss, and scattering, the use of directional antennas and beamforming techniques, and the impact of misalignment can also be modeled by stochastic geometry.

A stochastic geometry approach for mean interference power and outage probability analysis is considered in \cite{Kokkoniemi7880713}, in the context of a dense THz network operating over $\unit[0.1-10]{THz}$. The researchers model the interference as a shot noise process, and they assume directional antennas. Furthermore, although high antenna gains result in a lower probability of interference in the THz band, the interference level is much higher when it occurs. Stochastic geometry is also used in \cite{Wu9024417} to derive the exact and approximate distributions of the received signal power and interference, respectively; semi-closed-form expressions are derived for the coverage probability and the average achievable rate. Eventually, heterogeneous networks (HetNets) comprising mmWave, THz, and optical wireless communications should be analyzed \cite{wang2021stochastic}. Integrating small cells operating at mmWave and THz frequencies is required to meet the ever-increasing demand for ultra-high data rates; retaining sub-$6$~GHz cells overcomes the limited coverage of THz communications. Densifying the network with THz base stations should increase the average rate significantly but not the coverage probability that decreases when THz small cells dominate the network.



The performance of THz systems is also studied in use-case-specific scenarios. For example, the reliability and latency of THz communications are studied in the context of wireless virtual reality in \cite{Chaccour8763780,chaccour2020can}. High reliability can be achieved with proper densification, which is illustrated by deriving a tractable expression for system reliability as a function of THz system parameters. In another example \cite{Chen8761352}, the channel capacity and reliability are studied for the special case of THz wireless networks-on-chip communications \cite{Abadal8939356,Chen8896901}, where performance can be enhanced by proper choice of silicon layers and their thickness. Furthermore, user- and network-centric metrics for THz information shower systems are evaluated in \cite{Petrov7511129}, where 95\% of traffic from long-range networks can be offloaded, initiating heavy-traffic THz information shower sessions.

THz signals are also being considered for communications at atmospheric altitudes (where the concentration of water vapor decreases) among drones \cite{Krishna9240057,Krishna9240057}, jets, unmanned aerial vehicles \cite{Wang9248588}, high-altitude platform systems \cite{Tekbiyik9269928}, and satellites \cite{saqlain2021channel,tekbiyik2020reconfigurable}. Calculating the absorption loss through the atmosphere at higher altitudes permits such applications. In \cite{saeed2020terahertz}, communication at $\unit[0.75-10]{THz}$ is demonstrated as more feasible at higher altitudes than sea-level, with reported usable bandwidths of $\unit[8.218]{THz}$, $\unit[9.142]{THz}$, and $\unit[9.25]{THz}$ over a distance of $\unit[2]{km}$. In \cite{saqlain2019capacity}, the capacity of optoelectronic THz Earth-satellite links is analyzed, where the claim is made that 10 Gbps per GHz can be supported. 

{Similarly, the use of the THz} band for simultaneously providing high data rates and wide coverage data streaming services to ground users from a set of hotspots mounted on flying drones is studied in \cite{moorthyflytera}. By solving an optimization problem for resource utilization, much higher throughput can be achieved in mobile environments than in static environments. Furthermore, a holistic investigation on THz-assisted vertical HetNets is conducted in \cite{Tekbiyik9269928}. The study comprises, in addition to terrestrial communication links, geostationary and low-earth orbit satellites, networked flying platforms, and {\emph{in-vivo}} nanonetworks; accurate channel modeling is critical for harmony across all these applications. Analyses on THz-band nanocommunications are also reported in \cite{Zhang9321977} and in particular for body-centric applications in \cite{elayan2020information,saeed2020body,Yang9170555}.


\subsection{Experimental Demonstrations}
\label{sec:demos_exp}

Several experimental demonstrations have been conducted to verify the corresponding channel models and predicted performance metrics. Experimental results for the first true-THz absorption-defined window above $\unit[1]{THz}$ ($\unit[1.02]{THz}$) are reported in \cite{Sen8815595}, where tens of Gbps are demonstrated in a multi-carrier (OFDM) system over sub-meter distances. A typical software-defined PHY layer transceiver system consists of frame generation, modulation, pulse shaping, pre-equalization, noise filtering, frame synchronization, post-equalization, and demodulation. Although pre-equalization accounts for the frequency-selective response of components, post-equalization mitigates ISI and the frequency-selective channel response. A correlator filter is used for frame synchronization \cite{Sen8815595}. 

{In \cite{Ntouni2020Testbed}, the researchers demonstrated} a THz LoS link with digital beamforming, at an operating frequency of $\unit[300]{GHz}$, a channel bandwidth of $\unit[1.5]{GHz}$, and a communication distance reaching $\unit[0.6]{m}$; the results confirm an $\unit[11]{Gbit/sec}$ data rate. Tbps speeds are demonstrated over THz LoS links in \cite{Khalid7511280}, where NLoS links through first-order reflections are also feasible in cases of signal obstruction. These observations inspire the development of MIMO mechanisms that exploit spatial diversity in transmission, reception, and reflection. Furthermore, in \cite{Nallappan8488685}, a live streaming demonstration of an uncompressed 4K video using a photonics-based THz communications system (below $\unit[200]{GHz}$) is reported, where error-free transmission is achieved at a distance of $\unit[1]{m}$. Similarly, THz signals propagating through practical outdoor weather conditions and subject to indoor surface reflections are studied in \cite{ma2019propagation}. According to \cite{singh2019reliable}, THz communications can support outdoor communications with proper planning. 


Open-source large-scale distributed testbeds are being developed to facilitate experimental research in the mmWave and THz bands, such as \emph{MillimeTera} \cite{polese2019millimetera}. Several experimental testbeds that demonstrate multi-Gbps THz links over several distances have been reported for both electronic (at 240 GHz \cite{kallfass201564} 300 GHz \cite{jastrow2010wireless}, 625 GHz \cite{moeller20112} and 667 GHz \cite{Deal8059083}) and photonic systems (near 300 GHz  \cite{Chinni8510070}). Other real-time testbeds are proposed in \cite{Zhang2019Digital} ($\unit[30]{Gbit/sec}$ at \unit[325]{GHz}) and in \cite{Castro2019Broadband} ($\unit[100]{Gbit/sec}$ at \unit[300]{GHz}). In \cite{sen2021versatile}, a versatile experimental testbed debunked the limited THz transmission distance myth by demonstrating a more than $\unit[1]{Km}$ distance at $\unit[100]{GHz}$. 
{Furthermore, THz-wireless fiber extenders are experimentally} demonstrated at $\unit[300]{GHz}$ in \cite{Castro9269933,Castro9166263}, and a spectral-efficient 64-QAM-OFDM THz link is demonstrated in \cite{hermelo2017spectral}. However, we still lack experimental testbeds for THz UM-MIMO systems \cite{Singh9014135}. Nevertheless, thousands of graphene-based antennas promise to be embedded in small arrays, and electronic solutions promise high degrees of integration, which motivates research into UM-MIMO. The challenge is how to operate such arrays.


\section{THz Modulation Schemes and Waveform Design}
\label{sec:modulation}

Designing efficient THz-specific waveforms and modulation schemes is crucial for unleashing the THz band's true powers. Specific waveform designs can mitigate the limitations in THz sources and receivers. In contrast, optimized, adaptive modulation schemes can fully exploit the available spectrum. The choice of modulation schemes provides a compromise between low-complexity and high-rate PHY layer configurations. 

Because carrier-based systems at higher frequencies tend to use larger spectrum bandwidths per channel, simple modulation schemes that require very low complexity digital demodulation are favored (binary phase-shift keying and amplitude-shift keying). Nevertheless, novel modulation schemes and optimized multi-carrier waveform designs should be tailored for specific THz communication use cases. Early results on IEEE 802.15.3d-compliant waveforms are documented in \cite{shehata2021ieee}, which proposes an optimal THz envelope waveform that maximizes the spectral radiation efficiency under stringent emission constraints.



\subsection{Single-Carrier versus OFDM}
\label{SC_OFDM}

The use of multi-carrier waveform designs in future THz-band communications systems remains a controversial topic. The limited multipath components at higher frequencies result in frequency-flat channels that favor low-complexity wideband SC systems \cite{Dore8464918,Saad8885529}. Because THz beams are narrow under high antenna gains, the corresponding delay spread is reduced (survival of a single path), and the channel should be flat. This may require a deviation from OFDM. 

{OFDM is challenging to implement} in the context of ultra-broadband and ultra-fast THz systems (complex transceivers with Tbps digital processors still do not exist). The strict synchronization at the THz band severely hinders OFDM deployment, where frequency synchronization requires sampling rates on the order of multi-giga- or tera-samples per second \cite{Yuan9139316}. Furthermore, the high peak-to-average power ratio (PAPR) requirements also render OFDM ineffective in the THz band. 
{The limitations of DACs and ADCs also} prevent the digital generation of multi-band orthogonal systems \cite{Hossain8761547}. Special care should be given to mitigate the resultant Doppler effect with OFDM. Under perfect frequency and time synchronization, the cyclic prefix length in OFDM is chosen to account for the delay spread in the system, and the OFDM symbol length is proportional to the inverse of the Doppler spread. Hence, at THz frequencies, the cyclic prefix is relatively larger for the same delay spread conditions. This problem is highlighted in \cite{You7913686} and complicates OFDM design. Prospect solutions for this problem in the literature include beam-based Doppler frequency compensation schemes \cite{Va7742901,Chizhik1343892}. Although variations of multi-carrier designs result in the use of distinct THz spectra, spectral efficiency need not be the primary concern in the presence of huge bandwidths, and baseband complexity constraints might be more critical.


SC modulation for above-$\unit[90]{GHz}$ is proposed as a spectral- and energy-efficient solution for Tbps wireless communications \cite{Dore8464918}. Nevertheless, because bandwidth is abundant in the THz band, non-overlapping and perhaps equally-spaced sub-windows are efficient, as confirmed in several photonic THz experiments \cite{Lin7116524,Lin7786122}. Non-overlapping windows can be understood as SC modulation with some form of carrier aggregation, which is much less complex than OFDM and would thus enable the use of high-frequency energy-efficient power amplifiers. Transmission can be conducted in parallel over these windows \cite{Han6884190}, where each carrier would occupy a small chunk of bandwidth that supports a lower data rate. This relaxes design requirements (simpler modulation and demodulation) and reduces energy consumption while retaining an overall high-rate THz system. These benefits are at the cost of operating multiple modulators in parallel, where a high-speed signal generator is required to switch between carriers. 

{SC transmission is further advocated} in \cite{Palicot8105038}. SC has been proposed in WiFi 802.11ad (WiGig) \cite{Nitsche6979964} for mmWave communications \cite{Buzzi8101540}. The SC waveform can be complemented with simple continuous phase modulation (CPM) schemes such as continuous phase modulated SC frequency division multiple access (CPM SC-FDMA) and constrained envelope CPM (ceCPM-SC)  \cite{luo2016preliminary}. SCs can even provide higher power amplifier output power than cyclic-prefix OFDM. Furthermore, SC transceivers are resilient to PN \cite{Bicais9052917,tervo20205g}, especially when combined with a PN-robust modulation scheme. Low-order modulations have low PAPR and are robust to PN. By tracking the phase of local oscillators at the transmitter and receiver (e.g., using time-domain phase tracking reference signals), SCs can accurately estimate the PN at low complexity.


Frequency selectivity, however, can still arise due to frequency-dependent molecular absorption losses, frequency-dependent receiver characteristics, and the existence of several multipath components in indoor sub-THz systems. The bandwidth of each THz transmission window (approximately $\unit[0.2]{THz}$ \cite{Lin7786122}) can be much larger than the coherence bandwidth, which at $\unit[0.3]{THz}$ can be as low as $\unit[1]{GHz}$ in a multipath scenario and can reach $\unit[60]{GHz}$ with directional antennas \cite{Han7321055}. Frequency selectivity is a function of the communication distance, pulse bandwidth, and center frequency\cite{Yuan9139316}. {A frequency-selective system can also exist} because of the behavior of the THz receivers \cite{Sen8815595,jornet2017temporal}. 

Therefore, as a well-understood technology, OFDM can still be used for THz communications \cite{hermelo2017spectral}. Candidate multi-carrier schemes include cyclic-prefix OFDM (CP-OFDM),  block-based discrete-Fourier-transform spread OFDM (DFT-s-OFDM) \cite{Sahin2016}, continuous offset QAM-based filter-bank multi-carrier (OQAM/FBMC) \cite{bellanger2010fbmc}, and orthogonal time-frequency modulation (OTFS) \cite{hadani2017orthogonal}. A study on the performance and complexity trade-offs of such schemes in both the sub-THz and THz bands still lacks in the literature. Target parameters include PAPR performance, the complexity of different linear equalization schemes, and the impact of subcarrier spacing on PN.

OQAM/FBMC is considered a promising candidate for future wireless networks and cognitive radio applications as an alternative to CP-OFDM. OQAM/FBMC features very low out-of-band emissions, with the same bit error rate as CP-OFDM, but with enhanced spectral efficiency because no cyclic prefix is needed. Moreover, OQAM/FBMC relaxes synchronization requirements and offers less sensitivity to Doppler effects due to the use of well-localized prototype filters in the time-frequency domain such as PHYDYAS \cite{viholainen2009deliverable}. In contrast, all these advantages reflect negatively on the complexity of the equalizer because CP is not used to reduce ISI.
The direct form representation of FBMC consists of OQAM pre-processing, synthesis filter bank, analysis filter bank, and OQAM post-processing. 

OTFS \cite{hadani2017orthogonal} is promising for THz mobility scenarios with Doppler shifts that severely affect the automatic frequency control range. OTFS is robust to doubly selective channels because it modulates information symbols in the delay-Doppler domain. Alternative THz multi-carrier modulations such as wavelength division multiplexing (WDM) and Nyquist WDM \cite{takiguchi2020flexible} are also being studied.

\subsection{Optimized Modulation Schemes}
\label{mod_opt}

Modulation schemes can be further optimized to use the fragmented THz bandwidth to mitigate the absorption effect and turn it into an advantage. Accordingly, and because the THz channel response is distance-dependent, distance-aware multi-carrier schemes that dynamically optimize transmission window allocations are proposed in \cite{Han6884190}. Such schemes achieve Tbps data rates, an order of magnitude higher than fixed-bandwidth modulation schemes, over medium-range communications ($\unit[10]{m}$). Nevertheless, such schemes come at the expense of a slightly increased complexity because they typically require a control unit, a multi-carrier modulator, and a feedback path. In \cite{KrishneGowda7324520}, a {parallel sequence spread spectrum} is proposed as an alternative to OFDM for THz communications; it achieves 100 Gbps with simple receiver architectures that can almost be implemented almost entirely in analog hardware. In \cite{Peruga9145132}, constrained PSK is introduced as an energy-efficient modulation scheme for sub-THz systems.

Many more resources, however, can be dynamically optimized. For example, frequency allocation per AE is optimized in \cite{zakrajsek2017design} to maximize capacity as a function of the number of frequencies and AEs, antenna and array gains, and beamsteering angles. Moreover, a pulse-based multi-wideband waveform design is optimized in \cite{Han7321055} to enable communication over long-distance networks by adapting the power allocation criteria over a variable number of frames. The design incorporates pseudo-random time-hopping sequences and polarity randomization and accounts for temporal broadening effects and delay spread; a communication range of $\unit[22.5]{m}$ and a data rate of 30 Gbps are reported. A single-user and multi-user distance-aware bandwidth-adaptive resource allocation solution is proposed in \cite{Han7490372} that supports a data rate of 100 Gbps over a $\unit[21]{m}$ distance. The unique relationship between distance and bandwidth is also exploited to enable a multi-carrier transmission in \cite{Han6998944}.

A hierarchical modulation scheme is proposed in \cite{Hossain8761547} for a single-transmitter multiple-receiver system that supports multiple data streams for various users at different distances, by adapting the modulation order and symbol time. These optimization problems are extended to cases of large densities and user mobility. In such scenarios, multi-user interference is unavoidable. In \cite{singh2019parameter}, the opportunistic use of resources under mobility is maximized to satisfy the constraints on humidity, distance, frequency bands, beamwidths, and antenna placement. Furthermore, a stochastic model of multi-user interference is proposed in \cite{Hossain8736035}, alongside modulation schemes that minimize the probability of collisions. THz OFDM adaptive distance- and bandwidth-dependent modulations are also considered in \cite{Boulogeorgos8445864}.

\subsection{Pulse-Based Modulation}
\label{pulse-based}

Although continuous carrier-based transmission can be supported in the sub-THz range when the constraint on size is relaxed \cite{koenig2013wireless,6248357Song}, carrier-based transmission is still challenging at true THz frequencies. For example, it is not easy to generate more than short high-frequency pulses of few milli-watts with graphene at room temperature. Nevertheless, with large bandwidths, a reduction in spectral efficiency is acceptable, and pulse-based modulations can be used. Pulse-based SC on-off keying modulation spread in time (TS-OOK) that exchanges hundreds of femtosecond-long pulses between nanodevices is proposed in \cite{Jornet6804405}. By assuming time-slotted operations with a time slot $\bar{T}$, the Gaussian pulse is expressed as $p(t) \!=\! a\exp\left(-(t-b)^2/2{\bar{T}_p}^2\right)$, where $a$, $b$, and $\bar{T}_p$ are the amplitude, center, and spread of the pulse, respectively (Gaussian pulse has a duration $\bar{T}_p<\bar{T}$). Conversely, the raised cosine pulse of the carrier-based system is expressed as $q(t) \!=\! \textrm{sinc}\left(t/\bar{T}\right) \left(\cos\left(\pi\bar{\alpha} t/\bar{T}\right)\right)/\left(1-\left(2\bar{\alpha} t/\bar{T}\right)^2\right),$
where $\bar{\alpha}$ is the roll-off factor ($0\!\leq\!\bar{\alpha}\!<\!1$).

Pulse-based THz communications can achieve a Tbps data rate in nanonetwork scenarios \cite{Jornet6804405}. They have also been used in ultra-wide-band impulse-radio systems \cite{843135Win} and free-space optics \cite{brown2006experimental}. However, pulse-based systems are power-limited, especially for extremely wideband signals. Wideband pulses are thus typically used to achieve low-power, compact, and low-complexity sub-band transmissions. Nevertheless, by jointly optimizing the modulation and power allocation iteratively, Tbps rates are demonstrated in indoor communication paradigms in \cite{Moldovan7511275} by assuming realistic transmit, receive, and equalization filters under practical error rate constraints. An SC pulse-based approach is first addressed by optimizing the choice of modulation scheme by assuming a very long dispersive channel impulse response that accounts for ISI. Then, frequency division over multiple orthogonal sub-bands is considered, alongside efficient power allocation, to minimize the loss in the rate that is caused by finite-alphabet modulations. 


\begin{figure}[t]
\centering
\includegraphics[width=3.5in]{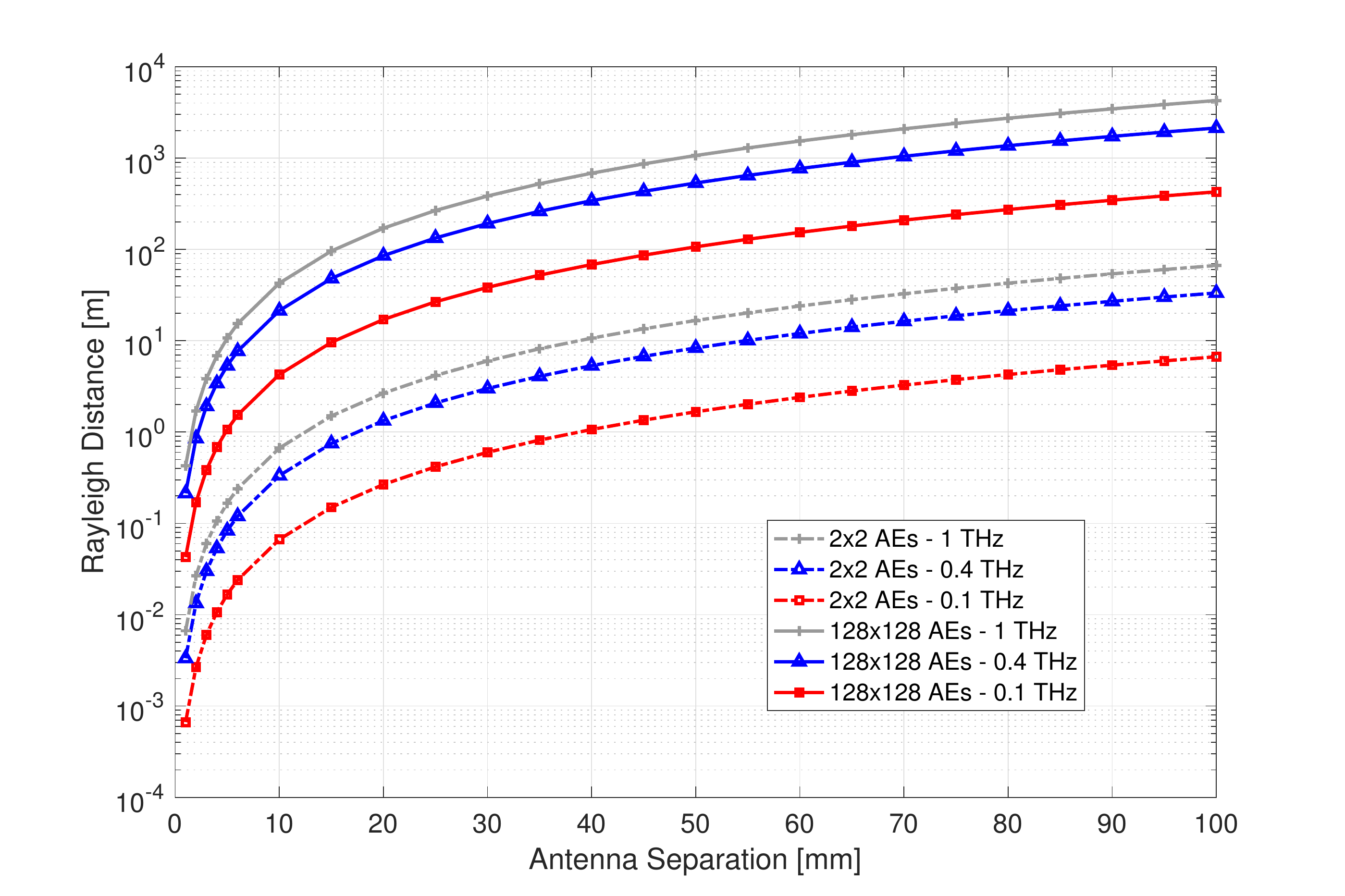}\vspace{2mm}
\caption{Achievable Rayleigh distances under different THz LOS configurations.}
\label{f:Rayleigh}
\end{figure}

\section{Reconfigurable UM-MIMO Arrays}
\label{sec:SPtuning}


Antenna gains and array gains are pivotal for overcoming path loss, where specific combinations of these gains are recommended for a specific communication distance and frequency of operation. Although a mmWave system typically requires a footprint of few square centimeters for several tens of antennas (not even sufficient to overcome the path loss over few tens of meters), a vast number of AEs can be embedded in a few square millimeters at THz frequencies. Furthermore, given the quasi-optical behavior under LoS dominance, a THz MIMO channel is sparse with multi-user beamforming and low-rank with spatial multiplexing. 

{As described in Sec. \ref{sec:sysmodel}, due} to high directivity and because beamforming is typically configured at the level of AEs within a SA, each SA is effectively detached from its neighboring SAs in a multi-user setting. Moreover, the role of baseband precoding reduces to defining the utilization of SAs, or to simply turning SAs on and off. However, in a point-to-point setup, the SA paths are highly correlated, and the channel is ill-conditioned. Nevertheless, good multiplexing gains are still achievable using sparse antenna arrays \cite{Torkildson6042312} that reduce spatial correlations in point-to-point LoS scenarios. The capacity of LoS MIMO uniform linear array channels is studied in \cite{do2020reconfigurable} over all possible antenna arrangements.



\begin{figure*}[t]
\centering
\includegraphics[width=4.5in]{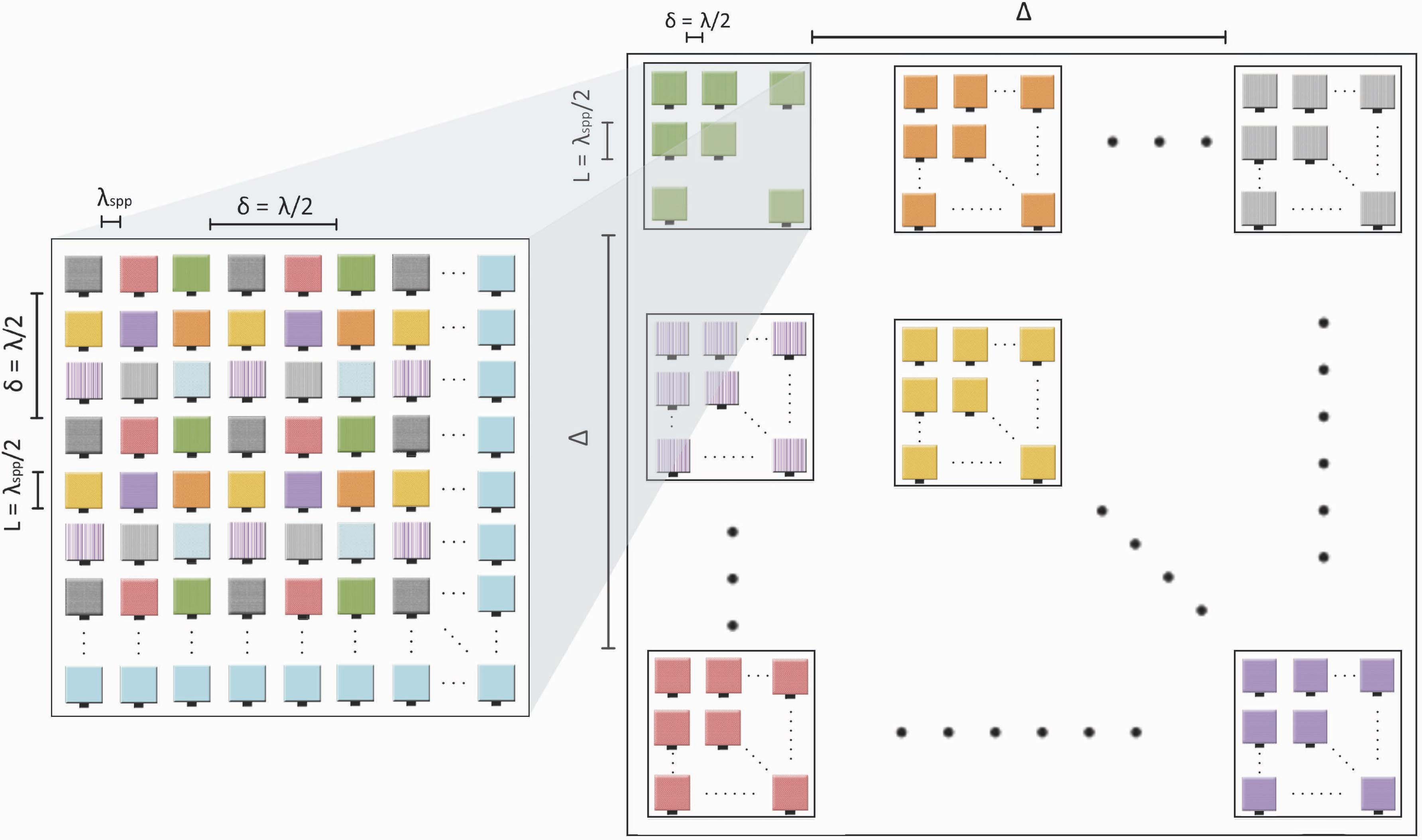}\vspace{2mm}
\caption{Interleaved antenna maps at the level of SAs and AEs (different colors denote different operating frequencies). }
\label{f:reconf}
\end{figure*}


\subsection{THz Spatial Tuning}
\label{sec:SPtuning2}



To enhance THz LoS channel conditions, spatial tuning techniques that optimize the separations between AEs can be applied \cite{Sarieddeen8765243}. Depending on the communication range ($d$) and the SA separation ($\Delta$), three modes of operation can be distinguished: (1) a mode where $\Delta$ is large enough so that the channel paths are independent and the channel is always well-conditioned, (2) a mode where $d$ is large compared to $\Delta$, and therefore the channel is ill-conditioned, and (3) a mode where $d$ is much larger than $\Delta$, larger than the Rayleigh distance, where the channel is highly correlated. If the communication range is less than the Rayleigh distance in the second mode, $\Delta$ can be adapted to enhance the channel conditions and achieve near-orthogonality \cite{Torkildson6042312,do2020reconfigurable}. Hence, the Rayleigh distance is an important metric for capturing THz system performance. For the number of transmit antennas $N$ and receive antennas $M$, this distance is expressed as \cite{6800118Wang}
\begin{equation}\label{eq:Rayleigh}
d_{\mathrm{Ray}} = \mathrm{max}\{M,N\} \Delta_r\Delta_t/\lambda,
\end{equation}
where $\lambda$ is the operating wavelength, and $\Delta_r$ and $\Delta_t$ are the uniform separation between SAs (also applicable to AEs) at the receiver and transmitter, respectively. 

{The LoS Rayleigh distances are illustrated} in Fig. \ref{f:Rayleigh}, as a function of array dimensions, operating frequencies, and antenna separations. Two AoSA configurations are simulated, $M\!=\!N\!=\!128\!\times\!128$ and $M\!=\!N\!=\!2\!\times\!2$ by assuming $\Delta_r\!=\!\Delta_t$. For small antenna separations, a large number of antennas are required to achieve multiplexing gains beyond several meters. For the same $\Delta$, higher frequencies and larger arrays result in larger Rayleigh distances. However, for the same footprint, increasing the number of antennas incurs a reduction in Rayleigh distance that is quadratic in $\Delta$. By finely tuning $\Delta$, multiple data streams over eigenchannels can be transmitted when \cite{Sarieddeen8765243}\vspace{2mm}
\begin{equation}
\Delta_\opt = \sqrt{\frac{zD\lambda}{M}},\vspace{2mm}
\end{equation}
for odd values of $z$. Such optimizations cannot be achieved in the third mode due to the limitation in physical array sizes. Alternative optimization schemes for antenna separations in LoS communications beyond the Rayleigh distance have been studied in \cite{6800118Wang,7546944Wang}. 

When combined with numerical optimization, we denote the corresponding adaptability in design by ``\emph{spatial tuning}.'' Spatial tuning is typically illustrated in the context of plasmonic antennas with a sufficiently large uniform sheet of AEs and where optimal configurations can be tuned in real-time \cite{Sarieddeen8765243,sarieddeen2021terahertz}. For example, a graphene-based sheet can consist of hundreds of uniformly-spaced active graphene elements mounted on a dielectric layer, which itself is mounted on a common metallic ground \cite{akyildiz2016realizing}. Hence, AEs can be contiguously placed over a 3D structure, and SAs can be virtually formed and adapted. For the desired communication range, a required number of AEs per SA is allocated. Then, the number of possible SA allocations, bounded by array dimensions and the number of RF chains, dictates the diversity/multiplexing gain.

Spatial tuning can be extended to include multi-carrier design constraints. For instance, nanoantenna spacings in plasmonic antenna arrays can be reduced to $\lambda_{\SPP}$ while still avoiding the effects of mutual coupling. Mutual coupling in graphene-based THz antenna arrays is studied in \cite{Zakrajsek7928818,Zhang8662671}. With the couple mode theory, the impact of mutual coupling on the response of nanoantennas is modeled via a coupling coefficient. Even at separations much less than $\lambda$, near field mutual coupling is negligible. This promising realization enables the practical implementation of compact THz systems in small footprints. In \cite{Zhang8662671}, the use of a frequency-selective surface structure that can be mounted between the array elements of an UM-MIMO array (behaving as a spatial filter) is proposed to reduce the mutual coupling effects further to negligible values. 

Placing AEs very close to each other, however, is not always beneficial because it reduces the spatial resolution and the achievable multiplexing gains. Furthermore, the inter-AE separation distance should not exceed $\lambda/2$, beyond which grating-lobe effects arise. By setting the separation between two active AEs to be $\lambda/2$, all the AEs in between would be idle. These antennas can be used for several purposes. Besides using them to increase the array gain, they can be configured to operate at different frequencies in a multi-carrier scheme that supports more users with the same array footprint, as illustrated in Fig. \ref{f:reconf}. Alternatively, neighboring AEs can be configured to operate at the same frequency in a spatial oversampling setup \cite{Rappaport8732419}, lowering the spatio-temporal frequency-domain region of support of plane waves \cite{Wang8470251}. The latter approach can be exploited for noise shaping, which results in a reduced noise figure and increased linearity.




\subsection{Index Modulation and Blind Parameter Estimation}
\label{sec:index_blind}



Spatial modulation (SM) schemes for THz communications are promoted in \cite{Sarieddeen8765243,saad2020generalized} as power- and spectrum-efficient solutions. The efficiency of SM at higher frequencies is highly dependent on the array design and channel conditions, as illustrated in \cite{Liu7547944,Sun8406871,Saad8880828} for mmWave systems. By mapping information bits to antenna locations adaptively, hierarchical SM solutions can be designed at the level of SAs or AEs. With SM, the number of bits that can be accommodated in a single channel use is
\begin{equation}\label{eq:nbofbits}
	N_b =  \underbrace{\log_2\left( M_tN_t \right)}_\text{SA} + \underbrace{\log_2\left( Q^2 \right)}_\text{AE} + \underbrace{\log_2\left( \abs{\mathcal{X}} \right)}_\text{symbols}.
\end{equation}  
The transmitted binary vector over one symbol duration can be expressed as $\mbf{b}\!=\![\mbf{b}_m \mbf{b}_q \mbf{b}_s]\!\in\!\{0,1\}^{N_b}$, where $\mbf{b}_m\!\in\!\{0,1\}^{\log_2\left( M_tN_t \right)}$ represent SA selection, $\mbf{b}_q\!\in\!\{0,1\}^{\log_2\left( Q^2 \right)}$ represent AE selection, and $\mbf{b}_s\!\in\!\{0,1\}^{\log_2\left( \abs{\mathcal{X}} \right)}$ correspond to the actual QAM symbol. The number of AEs and SAs and the constellation size can be tuned for the desired bit rate in such a design. By enabling the selection of various combinations of antennas simultaneously, a generalized index modulation scheme can be defined \cite{Younis5757786,Datta6554991}. Typical massive MIMO SM and generalized SM solutions \cite{Basnayaka7248610,Zuo8171123} should be revisited in the ultra-massive THz context. Furthermore, when enabling adaptive antenna-frequency maps, generic index modulation (IM) solutions that take full advantage of the available resources can be configured, in which information bits are also mapped to frequency allocations. The number of bits per channel use with IM can increase to
\begin{equation}\label{eq:nbofbitsGIM}
	N_b^{(\mathrm{IM})} =  \log_2 \left \lfloor {\bar{F} \choose F} \right \rfloor+ \log_2 \left \lfloor {S \choose M^2Q^2} \right \rfloor + \log_2\left( \abs{\mathcal{X}} \right),
\end{equation}  
where $F$ is the total number of available narrow frequency bands, and $\bar{F}$ and $S$ are the numbers of frequencies that can be supported and antennas that can be activated at a specific time, respectively. 

{Even more generalized} IM schemes can be achieved by jointly designing the spatial and frequency bit maps. Such designs could confirm to be particularly efficient in the THz band because a considerable number of AEs can fit in small footprints. The fragmented nature of the THz spectrum enables allocating multiple absorption-free spectral windows concurrently. However, the efficiency of these adaptive schemes is limited by the speed at which frequency hops can be executed. 
{Nevertheless, changing the frequency} of operation in the THz band can be achieved quickly without changing the physical dimensions of the transmitting antennas. Simple material doping or electrostatic bias change the Fermi energy of graphene, which dictates the frequency of operation. Software-defined plasmonic metamaterials are also a candidate solution, primarily for frequencies below $\unit[1]{THz}$. For SC systems with IM, constant-envelope modulations such as CPM are power-efficient \cite{Saad8885529}.

Such design compactness and flexibility can be further enhanced when complemented by THz-specific signal processing techniques at the receiver side. Instead of communicating transmission parameters with the receiver, blind parameter estimation can be conducted. For instance, in \cite{loukil2019terahertz}, a tertiary hypothesis test based on power comparison for THz antenna index and modulation mode detection is proposed and analyzed, alongside low-complexity frequency index detectors and modulation type estimators. Information bits can be assigned for the choice of modulation type as well. Although modulation classification \cite{Dobre,Sarieddeen7418322} is a classical signal processing problem, its applicability to the THz-band \cite{8540324Iqbal} can be particularly beneficial. Given the enormous possibilities of map-bit combinations, compressed sensing and machine learning techniques can be applied for detection and estimation.

\begin{figure*}[t]
\centering
\includegraphics[width=6in]{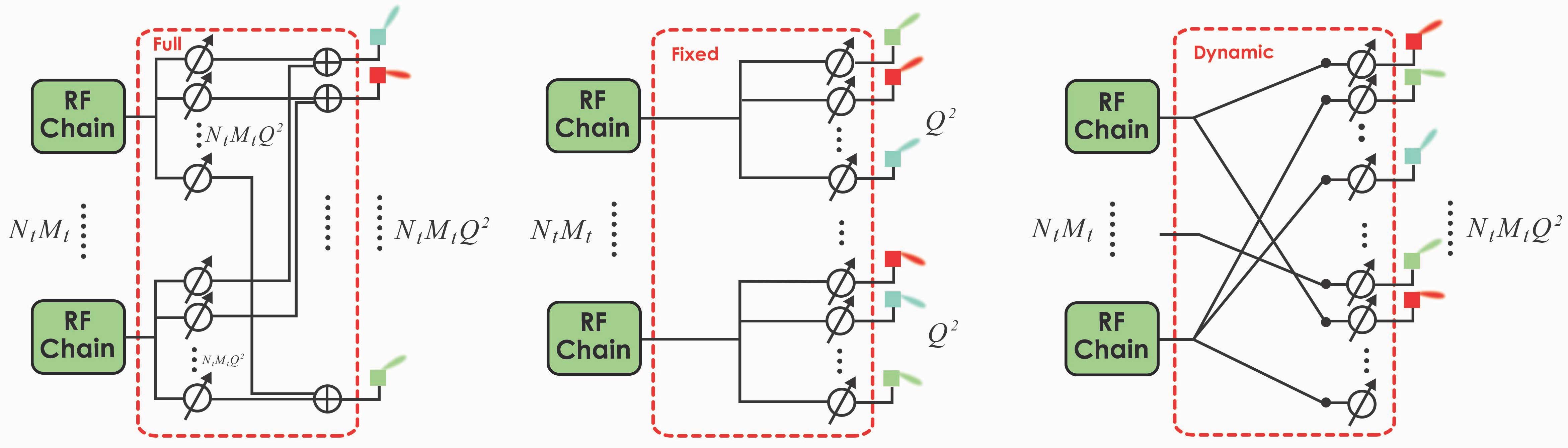}\vspace{2mm}
\caption{High-frequency hybrid beamforming architectures: Fully-connected, fixed AoSAs, and dynamic switching.}
\label{f:switching}
\end{figure*}

\section{Beamforming and Precoding}
\label{sec:beamforming}


As previously discussed, beamforming and precoding are critical to overcome the high path losses at high frequencies and exploit the THz channel's distance- and frequency-dependent characteristics. Beamforming enhances power versus distance such that the transmit power of CMOS THz devices can suffice for communication purposes. Because maintaining alignment is much harder at THz frequencies due to collimation, THz beamforming schemes should be fast. By the time 6G matures, fully digital arrays operating at mmWave frequencies and below will likely be readily available and capable of achieving near-optimal beamforming performance. However, it is unlikely that this would be the case at sub-THz and THz frequencies. 

{The motivation for hybrid beamforming} in the THz realm is similar to that in the mmWave realm \cite{Heath7400949}--there is a bottleneck in achieving prohibitively-complex and high-power-consuming fully-digital arrays (as highlighted in Sec. \ref{Sec:intro_devices}, power consumption remains a significant hurdle for the practical deployment of THz systems). THz hybrid beamforming is described in \cite{Lin7786122}, where a distinction is made between a fully-connected configuration, in which one RF chain drives the entire antenna array, and a configuration in which an RF chain drives a disjoint subset of antennas with a phase shifter per antenna (Fig. \ref{f:switching}). Due to limited hardware and processing capabilities, a single RF chain is typically assumed to drive an antenna array at the receiver side.


\subsection{THz Hybrid Beamforming}
\label{sec:hybrid}

Despite the limited power constraints of THz sources, the fully-connected configuration is expected to be power-aggressive, where the corresponding number of power-consuming combiners and phase shifters is very high \cite{skrimponis2020power}. Nevertheless, efficient fully-connected THz-band hybrid beamforming schemes can still be achieved \cite{Yan8873739,Yan9083795}. The most popular THz beamforming and precoding designs follow the AoSA configuration of Sec. \ref{sec:sysmodel}. In such architectures, analog beamforming is configured using a large number of AEs per SA to achieve spatial energy focusing. A beamsteering codebook design can be used per RF chain because THz phase shifters can be digitally controlled \cite{Chen6311430}. This can be implemented with beam scanning, ensuring that the received signal power is highest for a specific user.

{Digital precoding at the level} of SAs can be used to combat multi-user interference or define the utilization of SAs when interference is negligible due to high directivity. The precoding problem is typically formulated as an optimization problem that minimizes the mean square error between the received signal and the transmitted symbols under the power constraint. Simple zero-forcing precoding at the baseband should be sufficient in several THz scenarios. However, in highly correlated point-to-point THz links, higher-performing efficient nonlinear precoding techniques are required, such as block multi-diagonalization \cite{Nishimoto7552716,Nakagawa8108657}. The energy efficiency of the AoSA configuration is higher than that of the fully-connected one \cite{Lin7436794}. This difference is further emphasized when considering the nonlinear system power consumption model and insertion losses.

Finding the optimal allocation of SAs to enhance spectral efficiency, i.e., hybrid precoding with dynamic antenna grouping, is an important problem in THz UM-MIMO beamforming. Accordingly, switches can be inserted in a dynamic AoSA (DAoSA) architecture to tune the connections between SAs and RF chains \cite{Park7880698,Hu8338393,Xue8761071,Zhang8052157}. This flexibility is optimally achieved in a fully-connected architecture, in which establishing fully-dynamic connections requires an exhaustive search over all possible connections between RF chains and SAs. 
{However, such fully-connected} systems' complexity and power consumption are prohibitive at high frequencies and dimensions (thousands of switches). 

For exploiting this trade-off between spectral efficiency and power consumption, near-optimal and low-complexity THz hybrid precoding algorithms are proposed in \cite{Yan8756936}. The design problem is divided into two sub-problems: hybrid DAoSA precoding problem and switch selection. The system model corresponding to hybrid precoding in DAoSA architectures is a modification of \eqref{eq:sysmodel}, where the precoding and combining matrices, $\mbf{W}^{\Hr}_t$ and $\mbf{W}^{\Hr}_r$, are decomposed into their digital and analog components as \vspace{2mm}
\begin{equation}\label{eq:sysmodel_hybrid}
	\mbf{y} = \sqrt{\dot{p}}\mbf{C}_D^{\Hr}\mbf{C}_A^{\Hr} \mbf{H}\mbf{P}_A\mbf{P}_D \mbf{x} + \mbf{C}_D^{\Hr}\mbf{C}_A^{\Hr} \mbf{n},
\end{equation} 
where $\dot{p}$ is the transmitting power and $\mbf{y}$ and $\mbf{x}$ here have the dimension $N_s\times1$, with $N_s$ being the actual number of data streams, $\mbf{H}\in \mathbb{C}^{M_rN_rQ^2 \times M_tN_tQ^2}$, and $\mbf{n} \in \mathbb{C}^{M_rN_rQ^2 \times 1}$. Furthermore, $\mbf{C}_A\in \mathbb{C}^{M_rN_rQ^2\times M_rN_r}$ and $\mbf{C}_D\in \mathbb{C}^{M_rN_r\times N_s}$ are the analog and digital combining matrices, and $\mbf{P}_A\in\mathbb{C}^{M_tN_tQ^2 \times M_tN_t}$ and $\mbf{P}_D\in\mathbb{C}^{M_tN_t\times N_s}$ are the analog and digital precoding matrices, respectively. The achievable rate of this system model (subject to optimization) can be expressed as \vspace{2mm}
\begin{equation}\label{eq:rate_hybrid}
	R = \log_2 \left(  \abs{  \mbf{I}_{M_rN_rQ^2} + \frac{\dot{p}}{N_s \sigma^2} \mbf{H}\mbf{P}_A\mbf{P}_D \left( \mbf{P}_A\mbf{P}_D \right)^{\Hr} \mbf{H}^{\Hr} } \right),
\end{equation} 
where $\abs{\cdot}$ is the matrix determinant operator. Another dynamic SA architecture is proposed in \cite{Yan9348113}, which analyzes both quantized- and fixed-phase shifters.

Hybrid beamforming often entails user grouping and SA selection. In \cite{Lin7116524}, a THz multi-carrier distance-dependent hybrid beamforming scheme is proposed, in which user grouping is achieved through analog beamforming, alongside digital beamforming, power allocation, and SA selection mechanisms. The proposed solution enables users of different user groups to share frequencies while avoiding interference in the analog domain. Users in the same group, however, are assigned orthogonal frequencies based on a distance-aware multi-carrier scheme. SAs are then assigned to the data streams of a user group in the digital domain. {The same authors address} the hybrid beamforming problem for THz indoor scenarios in \cite{Lin7036065}.  A bound on the ergodic capacity is derived, and the impact of random phase shifter errors is analyzed. The relation between the required size and number of SAs and the communication distance is also established. The spectral efficiency gap between hybrid and digital precoding is smaller when the channel is sparser \cite{Ayach6717211}, making hybrid schemes suitable for multipath-limited THz channels. In \cite{yan2021joint}, joint inter- and intra-multiplexing and hybrid beamforming are proposed for widely-spaced multi-SA THz systems.


Developing wideband hybrid beamforming schemes for THz communications is also essential. For instance, in \cite{Yuan8647835}, the researchers propose an OFDM-based normalized codebook search algorithm for beamsteering and beamforming in the analog domain and a regularized channel inversion method for precoding in the digital domain. Two digital beamformers are used in a three-stage scheme to account for the loss in performance due to hardware constraints and the difference between subcarriers.
{ Similarly, in \cite{You7913686},} a beam division multiple access scheme is proposed for wideband massive MIMO sub-THz systems, which schedules a mutually non-overlapping subset of beams for each user. The algorithm is based on per-beam synchronization in time and frequency, considering the delay and Doppler frequency spreads, the latter of which are orders-of-magnitude larger at THz frequencies. An inaccurate narrowband assumption, in which the precoding and combining matrices of a wideband system are designed for a specific carrier frequency, induces the effect of beam split in THz communications. This phenomenon is mitigated in \cite{Tan9014304} using a THz-specific delay-phase controlled precoding mechanism, in which time-delay components are introduced between the RF chains and the phase shifters. The addition of such components creates frequency-dependent beams uniformly aligned with the spatial directions over the entire bandwidth.




%



In addition to the considerations above for beamforming in the THz band, all of which are extensions to similar considerations in the mmWave band, recent studies consider novel THz-specific beamforming schemes based on novel THz circuitry. For instance, in \cite{Andrello8645417}, a graphene-based dense antenna array architecture is proposed, in which each element is integrated by a THz plasmonic source, direct signal modulator, and nanoantenna. For such an architecture, novel dynamic beamforming schemes at the level of single elements and the level of the integrated array are proposed, where full phase and amplitude weight control can be achieved by tuning the Fermi energy of the modulator and AE. The researchers propose a codebook design for Fermi energy tuning that results in reasonably accurate beamforming and beamsteering; the power-density of the array increases non-linearly with its size. 
{Controllable THz frequency-dependent} phase shifters can also be achieved via low-loss integrally gated transmission lines \cite{Chen6311430}, the length of which determines the signal travel time, and hence the phase shift. Furthermore, graphene/liquid crystals have been proposed for magnet- or voltage-controlled THz phase shifters \cite{chen2004magnetically,wu2013graphene}. Because they are digitally controlled, such phase shifters only generate quantized angles.


Several other beamforming considerations need to be considered for future THz networks. For example, in a cell-free massive MIMO scenario \cite{zhang2019multiple}, distributed access points can each provide an excess of $100$ GHz to a user, especially under low mobility. Dense deployment of access points can guarantee short THz communication distances, even under significant blockage. Mobility and blockage are addressed in \cite{you2020network} in the context of network-massive MIMO scenarios in the mmWave and THz bands, where per-beam synchronization is proposed to mitigate the channel Doppler and delay dispersion; precoding beam-domain power allocation is reduced to a network sum-rate maximization problem. In other studies with system-level considerations, distance-aware multi-carrier hybrid beamforming based on beam division multiple access is proposed in \cite{Yuan9120579} to enable massive micro-scale THz networks; relay-assisted THz hybrid precoding designs are proposed in \cite{Mir9294139}.


\subsection{One-Bit Precoding}
\label{sec:one_bit}

The circuit power consumption, hardware complexity, and system cost significantly increase at higher frequencies and ultra-massive dimensions. The dominant sources of power consumption are ADCs in the uplink and DACs in the downlink. The power dissipation in converters scales exponentially in the number of resolution bits, and state-of-the-art DACs and ADCs can only achieve 100 gigasamples-per-second rates \cite{Laperle6616560}. Furthermore, the capacity requirements on the fronthaul interconnect links are also severe in large MIMO systems. Jointly reducing system costs, power consumption, and interconnect bandwidth with minimal performance degradation remains a challenge. 

As an alternative to reducing the number of converters using hybrid beamforming, the bit resolutions can be reduced through coarse quantization. This approach has the extra benefit of lowering the linearity and noise requirements, which is crucial in THz settings. In the extreme case of one-bit quantization~\cite{Mo_2014}, only simple comparators are required; automatic gain control circuits are no longer required. For high-amplitude resolutions, the power consumption of ADCs grows quadratically with the sampling rate. A one-bit quantization solution is proposed in \cite{neuhaussub} for sub-THz wideband systems, where the amplitude resolution is reduced while accounting for that by temporal oversampling. 

By modifying our system model from \eqref{eq:sysmodel}, the precoded transmitted symbol vector can be expressed as $\bar{\mbf{x}}=[\bar{x}_{1}\cdots \bar{x}_{b}\cdots \bar{x}_{M_tN_t}^{}]^{T}\!\in\!\bar{\mathcal{X}}^{M_tN_t\times1}$, where under finite-precision, the $\nth{b}$ symbol of $\bar{\mbf{x}}$, $\bar{x}_b\!=\! l_{\mathrm{R}}+jl_{\mathrm{I}}\in \bar{\mathcal{X}}$, has quantized in-phase and quadrature components, i.e., $l_{\mathrm{R}},l_{\mathrm{I}}\!\in\! \mathcal{L}$, where $\mathcal{L}=\{l_{0},l_{1},\cdots,l_{L-1}\}$ is the set of possible quantization labels and $\bar{\mathcal{X}}\!=\!\mathcal{L}\times\mathcal{L}$. For 1-bit quantization, we use $L\!=\!\abs{\mathcal{L}}\!=\!2$. Prior to precoding, the symbol vector $\mbf{x}$ is obtained by mapping the information bits to the original constellation $\mathcal{X}$. The base station then uses the knowledge of $\mbf{H}$ to precode $\mbf{x}$ into $\bar{\mbf{x}}$. $\mbf{x}$ and $\bar{\mbf{x}}$ need not be of the same size. With coarse quantization, an additional distortion factor exists due to finite precoder outputs. Because optimal precoding is exhaustive due to the cardinality of $\bar{\mathcal{X}}^{M_tN_t}$ in THz UM-MIMO systems, only linear quantized precoders \cite{Saxena7946265}, or perhaps very few optimized low-complexity non-linear quantized precoders \cite{Jacobsson7967843}, are feasible. Analyzing the system performance under quantization is typically conducted using the Bussgang decomposition \cite{tuugfe2020bussgang}.

The performance of THz indoor one-bit distance-aware multi-carrier systems is investigated in \cite{Li8648135} for a hybrid precoding AoSA architecture. The achievable rate is insensitive to changes in transmit power, and single-user transmission is robust to the phase uncertainties in large antenna arrays. The optimal beamsteering phase shifter direction is that of the LoS path. Furthermore, efficient modulation schemes can complement one-bit configurations. For instance, the zero-crossing modulation scheme \cite{fettweis2019zero} can mitigate THz impairments and relax hardware requirements using temporal oversampling and one-bit quantization at both the transmitter and receiver.

\subsection{THz NOMA}
\label{sec:NOMA}

Non-orthogonal multiple access (NOMA) techniques \cite{Dai8357810,Ding7973146} have been recently proposed to combat the loss of spectral efficiency in orthogonal multiple access schemes, especially when the resources are allocated to users with poor channel conditions. However, the lack of spectral efficiency is not a primary bottleneck for THz communications, given the readily available bandwidths that higher spatial resolution in beamforming limits the need for multiple access schemes. However, any additional spectral efficiency enhancement technique is welcome if its additional complexity cost is limited. 

{NOMA at higher frequencies} \cite{Zhu8798636} is thus more likely to be conducted over point-to-point doubly-massive MIMO links. In such scenarios, the concept of multiple access reduces to superposition coding of multiple data streams over a single link \cite{sarieddeen2021terahertz}. Nevertheless, calling the resultant configuration NOMA is not a misnomer because each THz beam can still be configured to serve multiple users. In this scenario, the role of NOMA can be essential in mitigating the hardware constraints in THz devices that limit the beamforming capabilities. 

For power-domain NOMA, and based on our system model, multiple data streams can be sent concurrently via superposition coding over different combinations of transmitting and receiving SAs that form overlapping effective channel matrices. Assume, for ease of construction, that the superposition of data symbols occurs at the lower layers of the MIMO channel matrix. Denote by $\mathcal{S}$ the set of power-domain multiplexed data streams of dimensions $S_i$, $i\!=\!{1,\cdots,\abs{\mathcal{S}}}$, such that $S_i\!\geq \!S_{i+1}$ and $S_1\!=\!M_tN_t$. The multiplexed transmitted symbol vector $\mbf{x}_i\!=\![x^i_{1}\cdots x^i_{n}\cdots x^i_{S_i}]\!\in\!\mathcal{X}^{S_i\times1}$ is allocated the contiguous set of SAs $N\!-\!S_i\!+\!1$ to $M_tN_t$. We thus use the effective channel matrices $\mbf{H}_{i}\!\in\!\mathbb{C}^{M_rN_r\times S_i}$, comprised of the columns $M_tN_t\!-\!S_i\!+\!1,M_tN_t\!-\!S_i\!+\!2,\cdots,M_tN_t$ of $\mbf{H}$. The equivalent baseband input-output system relation can then be expressed as
\begin{equation}\label{eq:sys_model_su}
\mbf{y}= \sum_{i=1}^{\abs{\mathcal{S}}} \mbf{H}_{i}\mbf{x}_i + \mbf{n},
\end{equation}
in which NOMA is achieved by assigning different power levels to the multiplexed transmitted symbol vectors. 

{For example, we can allocate a higher} power level $p_i$ to the symbol vectors $i$ of smaller dimensions, i.e., $p_i\!<\!p_{i+1}$. Each symbol $x^i_{n}$ thus belongs to a scaled complex constellation $\mathcal{X}^i$ ($\mathsf{E}[x_{n}^{i\Hr}x_{n}^i]\!=\!p_i$), and we use $\mbf{x}_i\!\in\!\tilde{\mathcal{X}}^i$, the lattice that includes all possible symbol vectors generated by $S_i$ $\mathcal{X}^i$ constellations. With higher degrees of reconfigurability in THz antenna arrays, generalized coding schemes encompassing antenna selection, frequency selection, and power allocation can be achieved. However, low-complexity detection and decoding schemes should complement such designs, particularly efficient successive interference cancellation at the receiver.

When sufficient multipath components exist, perhaps in scenarios where lower antenna gains are required (indoor sub-THz scenarios, for example), conventional single-cell multi-user MIMO-NOMA settings can still be achieved. Assume that the cellular users are divided into two groups, where the first group of users is uniformly distributed in an inner disk (${C}_1$) centered at the base station and of radius $ R_{\mathrm{N}}$, and the second group of users is uniformly distributed in an outer disk (${C}_2$) from $ R_{\mathrm{N}}$ to $ R_{\mathrm{C}}$. A base station with $N$ transmitting antennas simultaneously services two users, user 1 with $M_{1}$ antennas in ${C}_1$ and user 2 with $M_{2}$ antennas ${C}_2$, in the same frequency and time slot via power-domain superposition coding. The received vectors $\mbf{y}_1$ and $\mbf{y}_2$ at user 1 and user 2 can be expressed as\vspace{2mm}
\begin{align}\label{eq:sys_model}\vspace{2mm}
\mbf{y}_1 &= \mbf{H}_{1}\mbf{x}_1 + \mbf{H}_{1}\mbf{x}_2 + \mbf{n}_1, \\
\mbf{y}_2 &= \mbf{H}_{2}\mbf{x}_1 + \mbf{H}_{2}\mbf{x}_2 + \mbf{n}_2,
\end{align}
where $\mbf{x}_1$ and $\mbf{x}_2$ are the power-multiplexed transmitted symbol vectors and $\mbf{H}_1$ and $\mbf{H}_{2}$ are the corresponding channel matrices. In this scenario, NOMA is achieved by clustering users from the inner disk $C_1$ with users from the outer disk $C_2$ and assigning different power levels to the multiplexed transmitted symbol vectors. 

Adaptive superposition coding and subspace detection are considered for THz NOMA in \cite{sarieddeen2021terahertz}. Energy efficiency in a THz MIMO-NOMA system is also addressed in \cite{zhang2020energy} by optimizing the user clustering, hybrid precoding, and power allocation mechanisms. Furthemore, energy-efficient resource allocation in downlink THz-NOMA systems is studied in \cite{Zhang9257475}. Although the attainable gains of power-domain MIMO-NOMA remain unclear, THz scenarios suggest a compelling use case. 

{In their recent study, \cite{clerckx2021noma},} the researchers argue that MIMO-NOMA can misuse the spatial dimension by incurring a multiplexing gain loss because of fully decoding streams in SIC. Such loss is particularly evident when comparing MIMO-NOMA to other MIMO schemes, such as conventional multi-user linear precoding (MU-LP) or rate splitting (RS). Nevertheless, compared to OMA, the benefits of NOMA are clear. Furthermore, the proposed efficient SIC using subspace detectors in \cite{sarieddeen2021terahertz} combats this multiplexing gain reduction. Spatial precoding in MU-LP fails to overcome inter-stream interference under ill-conditioned near-singular THz channels matrices. Accordingly, power-domain NOMA is a vital enabler for THz data multiplexing. 

\section{THz Baseband Signal Processing}
\label{sec:baseband}

Efficient baseband signal processing is critical for mitigating the impairments of novel THz-band devices and enabling operations beyond 100 Gbps \cite{Rodriguez9083827}. The true bottleneck at the baseband is the lack of energy-efficient transceivers that can approach a Tbps data rate \cite{Weithoffer8109974}, where the sampling frequency is still on the order of {tens} of gigasamples-per-second in state-of-the-art ADCs and DACs. Efficient signal processing across all baseband blocks is required to fill this gap. Due to Moore's Law's diminishing effect, limited advances in chip power density and baseband computations are expected from silicon scaling. Therefore, incorporating application-specific integrated circuit (ASIC) architectures in a holistic framework is vital for achieving Tbps data rates and decreasing the time-to-market of THz-operating equipment. Joint algorithm and architecture optimization across all baseband blocks optimizes latency, area efficiency, power consumption, and overall throughput. 

Although channel code decoding dominates baseband computations \cite{Kestel8625324}, the entire baseband chain should be optimized. Accordingly, algorithm and architecture co-optimization for synchronization, data decoding and detection, and channel estimation is required. Such optimization can exploit the specific THz propagation characteristics, such as angular and temporal sparsity. Implementing THz-band high-resolution precoding and beamforming algorithms in the baseband's digital domain is challenging due to massive dimensionalities and the lack of processing capabilities. Such holistic solutions are especially crucial for massive antenna dimensions and high mobility wideband scenarios. We highlight recent advances in low-complexity channel estimation, channel coding, and data detection schemes in the following sub-sections. 


\subsection{Channel Estimation}
\label{ch_est}





Channel estimation in the THz band is very challenging in mobile scenarios, where accurate CSI is required in beamforming mechanisms and for accurately directing beams to avoid misalignment issues. Accurate CSI is essential in the absence of an LoS path. Furthermore, frequent channel estimation might also be required for fixed LoS point-to-point THz links because, at the micrometer wavelength scale, slight variations in the environment can introduce significant channel estimation errors. Classical channel estimation techniques should also be revisited by considering low-resolution quantization and hybrid analog and digital designs. Several techniques can be considered to reduce the complexity of THz-band channel estimation, such as fast channel tracking algorithms, lower-frequency channel approximations (exploiting outband signals), compressive-sensing-based techniques, and learning-based techniques, to name several.

Compressive-sensing techniques for sparse channel recovery in THz channel estimation \cite{Schram8645479} are inspired by their successful use in mmWave communications \cite{Alkhateeb6979963}, where channels are even less sparse. These approaches can exploit sparsity in the dictionary, delay, or beamspace domains. Researchers investigate variations of greedy compressive sampling matching pursuit \cite{Schram8645479}, orthogonal matching pursuit (OMP) \cite{singh2018compressed}, and the least absolute shrinkage and selection operator (LASSO) \cite{ghods2019beaches}. 
{Furthermore, in \cite{schram2019approximate}, approximate} message passing (based on belief propagation in graphical models) and iterative hard-thresholding are argued to be an efficient compressed-sensing-based technique for THz channel estimation. Nevertheless, learning-based THz channel estimation schemes are most efficient at higher dimensionalities \cite{he2020beamspace}. Deep kernel learning based on the Gaussian process regression is explored in \cite{Nie9024624} for multi-user channel estimation in UM-MIMO systems over $\unit[0.06-10]{THz}$.


Despite channel sparsity, the real-time THz channel estimation complexity overhead can be significant in a dense multi-user wideband scenario with many paths; a large number of measurements might be required for compressive-sensing-based estimation. Accordingly, traditional minimum mean square error (MMSE) and least square channel estimation methods can be used to estimate the second-order statistics of THz channels \cite{Lin7786122}. 
{Furthermore, joint activity detection} and channel estimation is an efficient technique to reduce the use of pilots and the complexity of computations in wideband random massive-access THz systems \cite{shao2020joint}. Fast channel tracking is an alternative approach to reduce the channel estimation overhead in high-mobility scenarios, as illustrated in \cite{Gao7582545} for THz beamspace massive MIMO. Eventually, THz channel estimation algorithms should be complemented by efficient hardware implementations to verify their practicality, as demonstrated in \cite{Mirfarshbafan9203990} for sparsity-exploiting mmWave beamspace channel estimation algorithms \cite{Mirfarshbafan9052952}.

\subsection{Channel Coding}
\label{ch_cod}

Data detection and decoding algorithms and architectures require low latency and high energy efficiency and throughput to bridge the Tbps gap in baseband signal processing. They should be highly parallelizable (spatial and functional parallelism) and should exhibit significant data locality and structural regularity. Channel coding is hitting the implementation wall because it is the most computationally demanding baseband process \cite{Kestel8625324}. The three central candidate coding schemes for 6G are Turbo, LDPC, and Polar codes. 
{Although Turbo and LDPC decoders} are both executed on data-flow graphs, Turbo decoding is inherently serial, and LDPC decoding is inherently parallel. In contrast, Polar decoding is typically performed on a tree structure and is inherently serial. Due to their parallel nature, LDPC decoders provide higher throughput \cite{Mansour1255474}. However, Polar and Turbo codes provide higher flexibility in code rates and block sizes \cite{Weithoffer8109974}, which is much required in 6G. 

A modular framework for generating and evaluating high-throughput Polar code decoders is presented in \cite{kestel2018polar}, where soft cancellation algorithms are explored. Turbo codes have advanced significantly towards beyond-100-Gbps operations \cite{Weithoffer8625377,Klaimi8625359,Le8960428}. Achieving a Tbps throughput with Polar codes is also addressed in \cite{Sural8880815}, where low-latency majority logic and low-complexity successive cancellation are combined for decoding, alongside an adaptive quantization scheme for log-likelihood-ratios (LLRs). This scheme achieves Tbps in a $\unit[7]{nm}$ technology implementation while occupying a $\unit[10]{mm^2}$ chip area and consuming $\unit[0.37]{W}$ of power. 
Furthermore, Polar codes complemented with guessing random additive noise decoding (GRAND) are shown to be computationally efficient for short-length, high-rate codes \cite{Duffy9086275,Duffy8630851}, which is promising for THz-based control channel communication, especially if further knowledge on THz system noise is developed.

Although such advances in coding schemes serve the ultimate goal of THz communications, which is achieving Tbps operations, they are blind to the inherent characteristics of THz channels. Nevertheless, the THz channel can be considered in MIMO detection schemes and joint modulation, coding, and detection algorithms and architectures.



\subsection{Data Detection}
\label{data_det}

Although channel code decoding is the most computationally demanding baseband processing block, data detection also adds a significant computational burden, especially in doubly-massive MIMO systems. Channel hardening occurs in conventional massive MIMO systems at lower frequencies, with many antennas at the base station and several antennas at the receiving equipment. With channel hardening, simple linear detection schemes such as zero-forcing and MMSE can achieve near-optimal performance \cite{Ngo_2013}. This is not the case in THz systems, where symmetric doubly-massive MIMO systems are common \cite{Buzzi8277180}, especially because compact large THz antenna arrays can be embedded in the UE. In the latter scenario, the channel tends to be highly correlated, especially under THz LoS-dominance. Inter-channel interference prohibits using simple linear detection schemes that fail to decouple spatial streams and result in noise amplification.

Consequently, more sophisticated non-linear detection schemes should be considered. However, the complexity of optimal non-linear detection schemes that achieve near-maximum likelihood performance is prohibitive at large dimensions. Therefore, novel THz-specific MIMO detectors that can achieve near-optimal performance with reasonable complexity are required. Conventional near-optimal detectors mainly replace the full-lattice search over all candidate transmit vectors in maximum likelihood detection with a reduced search over a reduced space of vectors closer to the truly transmitted vector (from a Hamming distance perspective). 
{Such reduced-complexity detectors are mainly} variations of sphere decoding schemes \cite{2014_sphereP1_mansour,Sarieddeen7511059}. Although sequential processing in sphere decoding results in variable complexity and limits parallelism, several algorithmic and architectural optimizations have been proposed \cite{2014_sphereP2_mansour} to fix its complexity \cite{barbero2008fixing}. In \cite{ltaief2019efficient,arfaoui2016efficient}, the complexity of sphere decoding is reduced by casting memory-bound computations into compute-bound operations, and real-time processing is maintained by using graphics processing units.

However, even fixed-complexity sphere decoding is prohibitively complex if used for UM-MIMO detection. Recently, several detection algorithms suitable for large doubly-massive MIMO systems have been proposed. Such algorithms are based primarily on local search criteria \cite{Li_2010}, heuristic tabu search algorithms \cite{Srinidhi_2011}, message passing on graphical models \cite{Narasimhan_2014}, Monte Carlo sampling \cite{Datta_2013}, and lattice reduction \cite{Zhou_LR_2013}. The Bell Laboratories Layered Space-Time (BLAST) detection algorithm is also modified to support ultra-high data rates in massive MIMO scenarios in\cite{Shental8167319}. Furthermore, perturbation-based regularizations can be used for equalization with ill-conditioned channels \cite{Ballal8070950}.

One family of detectors in particular that can achieve an acceptable trade-off between performance and complexity in large highly-correlated MIMO channels is the family of subspace detectors \cite{Sarieddeen8186206,sarieddeen2018high,Sarieddeen7471826,Sarieddeen7565040,Sarieddeen8292405,Mansour9298955}. 
Subspace detectors mainly exploit channel puncturing to reduce complexity and enhance parallelism. Because the computational cost of MIMO detectors is proportional to the number of nonzero elements in a channel matrix (most detectors involve back-substitution and slicing operations), by puncturing the channel into a specific structure, the detection process can be simplified and accelerated. 
{More importantly, subspace detectors} can break the interconnection between spatial streams, significantly enhancing parallelism at a marginal cost of multiple channel decompositions. Channel puncturing can be generalized to channel shortening, which can mitigate ISI in SC THz systems \cite{Hu7883887}. The performance gap between optimal channel shortening (from an information-theoretic perspective) and channel puncturing can be covered by adding MMSE
prefilter and channel-gain compensation stages \cite{mansour2020optimal,Sarieddeen8437872}. Subspace-marginalized belief propagation can also be adopted for high-frequency data detection \cite{Takahashi9148674}.


Other THz-specific data detectors include envelope- and energy-based detectors \cite{Bicais9217341}, which enable direct baseband operations without frequency down-conversion, bypassing phase impairments and enabling non-coherent detection that is inherently robust to PN. 
Low-complexity energy detection is studied in \cite{Sharma9293782} for pulse-based systems. Compressed detection with orthogonal matching pursuit is also considered for sparse pulse-based multipath THz communications \cite{singh2018compressed}.

MMSE precoding and detection are explored in \cite{Peng8599164} by assuming sparse channel matrices for a broadband SC THz system. Finite-alphabet equalization can also confirm helpful in THz scenarios. Coarsely quantizing the equalization matrix reduces the complexity, power consumption, and circuit area \cite{9053097}, increasing the speed of most critical baseband tasks, such as downlink precoding and data detection. Such coarse equalization enables all-digital massive multi-user MIMO equalization in mmWave systems \cite{Castaneda9110827} and should scale up efficiently with THz transceivers. Indoor THz communications Tbps rates under finite alphabets are demonstrated in \cite{Moldovan7511275}, where a frequency-division scheme of multiple sub-bands is used to relax the requirements on ADCs and DACs.



\subsection{Joint Coding, Modulation, and Detection}
\label{joint_cod_det}


One method of incorporating the effect of the THz channel in decoding is to consider iterative detection and decoding schemes. If MIMO detectors generate soft-output LLRs, these LLRs can be fed as soft inputs to the decoder, whose output LLRs can then be fed again as soft inputs to the detector \cite{tomasoni2012hardware}. However, the extra complexity in iterations and computing soft-output values in the detector should be considered. In the particular case of MIMO detection with Polar code decoding, iterations can be configured per stream detection after each successive decoding cancellation; the output of every step in the decoder can thus be used to enhance channel equalization.

Parallelizable detectors are favored for such designs. However, with parallelizability, the transmission vectors per stream would typically consist of a smaller number of bits. Although Polar and Turbo decoders can cope with this, LDPC decoders might not perform well. Larger modulation types can be considered to increase the number of bits per stream for enhanced decoding, 1024 QAM and beyond \cite{Sarieddeen7418324}, for example. However, THz systems do not perform well with higher-order modulations due to the increased complexity and the PN effect. Alternatively, deeply-pipelined MIMO architectures can be used to aggregate data for the decoders, but this comes at the expense of reduced throughput. Nevertheless, because different transmission vectors within a decoder block can be independent, multiple detectors can operate in parallel.


In addition to joint channel coding and data detection, joint modulation design and coding can use resources most efficiently, especially in multi-wideband THz communications.  Efficient probabilistic signal shaping techniques \cite{Bocherer7307154,Icscan8636931,elzanaty2020adaptive} can be used accordingly, especially in highly reconfigurable UM-MIMO systems. In \cite{Jornet5984951}, the researchers demonstrate that due to the peculiarity of noise in the THz band, proper use of channel codes can increase single-user and network capacity beyond classical networks with AWGN. 

Furthermore, THz-specific coding schemes are also being introduced at the network level. For instance, systematic random linear network coding (sRLNC) is proposed in \cite{phung2020error,phung2020improving} for generic THz systems, in which coded low rate channels carry redundant information from parallel high rate channels. By tuning the transmission and code rates, the number of channels, and the modulation format, fault-tolerant high-throughput THz communications can be supported at different communication ranges. Furthermore, in the context of THz index modulation, joint data detection and parameter estimation can be executed at the receiver side \cite{Sarieddeen7472369}. In the particular case of subspace detection, the entire process can be parallelized \cite{Sarieddeen7588171,Sarieddeen7564725}.

Finally, THz digital baseband operations of any complexity might be prohibitive in some use cases. Therefore, designing an all-analog THz baseband chain is a reasonable solution. Such design should include, in addition to analog modulation and detection schemes, all-analog decoders \cite{Hagenauer708738}, which are based on soft-output transmission and detection (bit LLRs represented by currents and voltages). All-analog MIMO schemes can be achieved using a continuous mapping scheme, which is noise-limited but might perform well under interference.

\section{Reconfigurable Surfaces}
\label{sec:surfaces}

\begin{figure*}[t]
\centering
\includegraphics[width=3.5in]{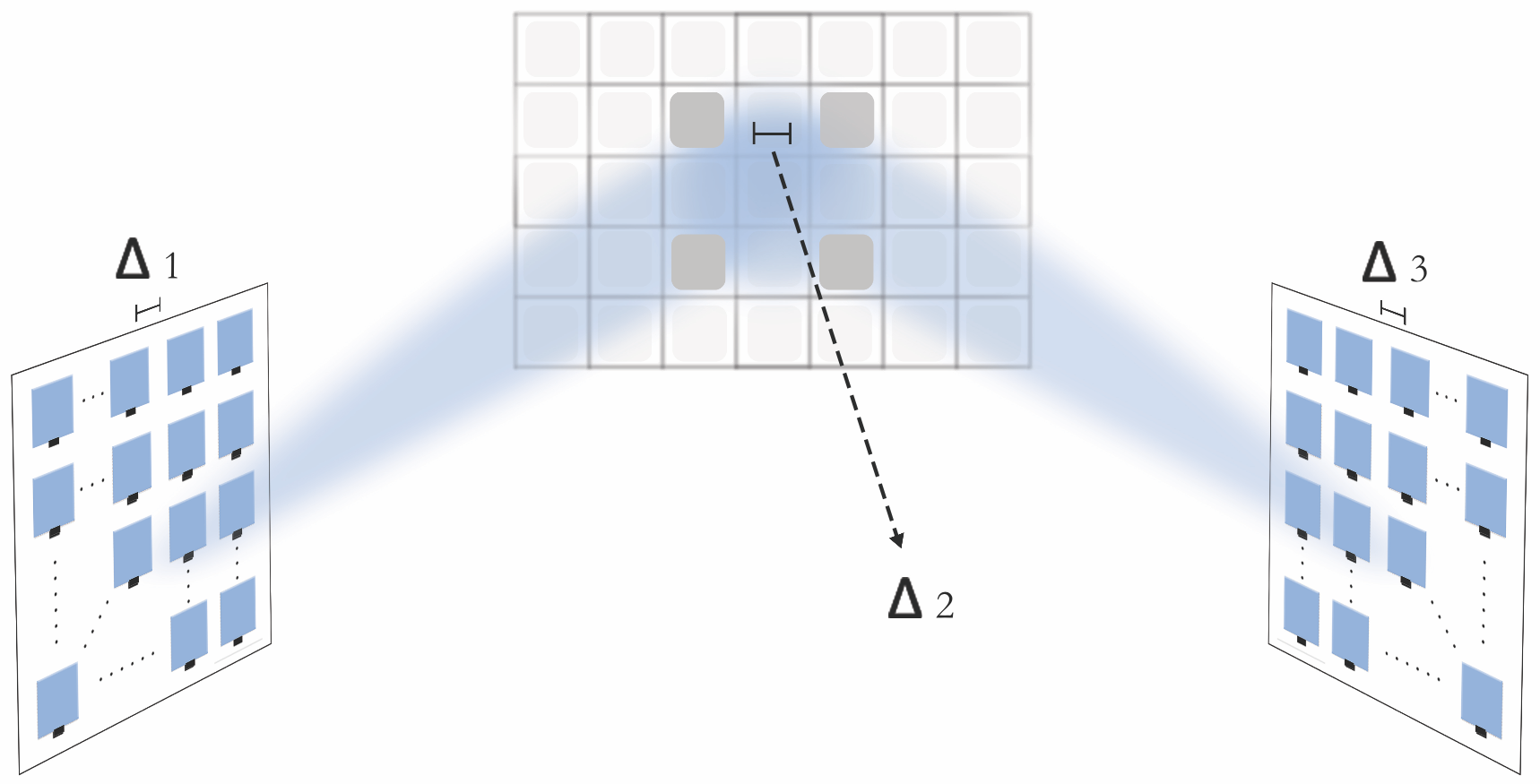}\vspace{2mm}
\caption{THz communications assisted by reconfigurable intelligent surfaces.}
\label{f:IRS}
\end{figure*}

\subsection{Emergence of IRS Technology}

One of the recent hot research topics in wireless communications is creating an intelligent, programmable environment for communication \cite{di2020smart}. A straightforward approach is to install large active arrays of AEs, also known as active large intelligent surfaces (LISs) \cite{Hu8319526,Hu8264743,Hu8647606} on indoor and outdoor walls and other structures. This approach is a special case of UM-MIMO and is suitable for THz scenarios. With minimal restrictions on how to spread antennas over a surface, the mutual coupling effects can be avoided, and channel correlation can be reduced in LoS environments (spatial tuning of Sec. \ref{sec:SPtuning2} can be easily configured). 
{Furthermore, channel estimation} and feedback mechanisms can be easily achieved in active LIS setups, essential for achieving low-latency THz communications. Another form of large surfaces is the concept of holographic MIMO surfaces \cite{huang2019holographic}. Because holographic MIMO borrows techniques from the optical domain, its implementation in the THz band should be more convenient than at lower frequencies \cite{Wan9374451}.


Passive IRSs \cite{ntontin2019reconfigurable,nadeem2019intelligent} are gaining most of the attention. IRSs are typically implemented using reflective arrays or software-defined metasurfaces, which introduce phase shifts at the level of reflecting elements (non-specular reflections) to focus and scale up the power of reflected signals and steer beams in a particular direction. These results can be achieved without requiring complex encoding and decoding schemes or additional RF operations \cite{Basar8796365}. 

{Phase shifts can also be} implicitly achieved by tuning the impedance or length of delay lines. Metasurface elements can be much smaller than those of reflectarrays (which typically follow the half-wavelength rule). Hence, they support more functionalities, such as polarization manipulation and absorption of incident waves. Reflections from tiny reflecting elements (of sub-wavelength size) form scattering in all directions, of which the combined effect is beamforming. The control complexity of metasurfaces, however, can be higher. Compared to LISs, both types of IRSs are passive (LIS and IRS are used interchangeably in the literature). Nevertheless, IRSs should be electronically active at some level, perhaps to transmit pilots and for operational purposes. 

IRS systems are particularly favorable in the THz band, where they can introduce controlled scattering to extend the very limited achievable communication distances and enable multicasting. Hence, IRSs can add synthetic multipath components to enhance the performance of multipath-limited THz systems. The necessity for IRSs operating at THz frequencies also arises from the fact that regular coherent large antenna arrays are not easily achieved with the very small size of AEs. Even the relaying technology at THz is not mature. IRS systems are thus a viable solution.

{Another argument for IRS THz deployments} stems from the limitations of surfaces themselves. For IRSs to achieve SNRs comparable to those of massive MIMO, or outperform a classical half-duplex relay system, a large number of reflecting elements is required. This could result in physically large arrays that are harder to deploy and subject to beam squinting. {The near-field behavior of IRSs is studied in \cite{bjornson2019demystifying,Bjornson9184098}, where a physically accurate near-field channel gain expression is derived for planar arrays, also considering the mismatch between the incident wave and the polarization of antennas.} 
{At THz frequencies, however, an electronically-large IRS} (compared to the operating wavelength) can be achieved in very small footprints, suggesting that dense THz-band IRS deployments for short-range communications can be easily achieved. In addition to regular IRS functionalities, at THz frequencies, reflectarrays can be used at the transmitter to generate and direct a THz beam excited by a close THz source \cite{Han7579223}. IRS signal processing enables both communications and sensing \cite{bjornson2021reconfigurable}.

\subsection{THz IRS Material Properties}

In addition to the THz-specific communications system considerations, several device and material properties favor THz-band IRS deployments, where the corresponding technologies might be mature before THz MIMO technology. Although THz-IRS CMOS technology \cite{Venkatesh2020,Liu2014} has low power consumption, it is limited by the maximum clock speed. Furthermore, CMOS integrates easily with different cell designs but is challenged by parasitic capacitance leakage. The CMOS unit cells of subwavelength dimensions can also result in large footprints. Micro-electro-mechanical systems (MEMS) for THz surface design \cite{Zhao2018,Han2017} are also limited by switch movement speed and control signaling, where each switch requires independent current and the resultant power consumption is relatively high. Because they are mechanical, MEMS are also more subject to faults and error and exhibit relatively big footprints. 

{Conversely, graphene-based technology} is low-power-consuming, easy to implement and integrated into small footprints, and has a simple biasing circuit. Nevertheless, graphene is limited by the voltage-implementing controller. Graphene-based metasurfaces can control the chemical potential of reflecting elements via electrostatic biasing, which varies the complex conductivity to achieve phase control \cite{lee2012switching}. Graphene-based digital metasurfaces combining reconfigurable and digital approaches are studied in \cite{Hosseininejad8745693}, where beamsteering is achieved by dynamically adjusting a phase gradient along the metasurface plane. A graphene-based metasurface is also proposed in \cite{dash2019wideband}, where a two-dimensional periodic array of graphene meta-atoms guarantees a wideband perfect-absorption polarization-insensitive reconfigurable behavior at THz frequencies.

THz metasurface-based beamsteering techniques can achieve wide-angle ranges at high compactness and low weight. In \cite{spada2019metasurface}, the proposed curvilinear THz metasurface design is argued to be independent of the geometry and the frequency. Metamaterials and metasurfaces operating in the THz band can provide flexibility for generating orbital angular momentum and polarization conversion \cite{fu2019terahertz}. 
{Furthermore, the concept of HyperSurfaces} is proposed in \cite{Liaskos8466374} for THz communications. HyperSurfaces are composed of a stack of virtual and physical components that can enforce lens effects and custom reflections per tile. 

In addition to graphene, thermally or electrically tunable vanadium dioxide and liquid crystals and micro-electromechanical systems have also been considered as candidates for efficient THz steering technologies \cite{fu2019terahertz}. Novel intelligent plasmonic antenna array designs for transmission, reception, reflection, and waveguiding of multipath THz signals are further studied in \cite{nie2019intelligent}, where an end-to-end physical model is developed. Nevertheless, future metasurfaces should support higher reconfigurability and sensing accuracy to support spatially-sensitive THz communications. A novel distributed control process, perhaps aided by optical internetworking, can guarantee fast adaptation \cite{Liaskos8466374}.

\subsection{Signal Processing for THz IRS}

Performance analysis studies for IRS-assisted communications include \cite{nadeem2019large,nadeem2019multi,bjornson2019demystifying,jung2019optimality,Hu8580675,Jung8948323,Ozdogan8936989,Bjornson8888223}. However, the corresponding performance limits at high frequencies \cite{ma2020intelligent,ozdogan2020using,ning2019channel,Tasolamprou8788546,Han7579223}, are still lacking. Most analytical studies consider lower-frequency scenarios and assume a downlink multi-user system model, in which an IRS is in the LoS of a base station assisting in reaching multiple users, each having a small number of antennas. For such setups, in \cite{nadeem2019large}, the minimum achievable signal-to-interference-plus-noise ratio is studied, both when the channel between the IRS and the base station is full-rank or rank-one. 

{The optimality of passive beamforming} in IRSs is studied in \cite{jung2019optimality}, where a novel modulation scheme that avoids interference with existing users is proposed, alongside a resource allocation algorithm. Opportunistic scheduling is further studied in \cite{nadeem2019multi} as a means to achieve good multi-user diversity gains in spatially correlated LoS scenarios. Furthermore, the effect of random blockages in large-scale surface deployments is studied in \cite{kishk2020exploiting} (blockage mitigation is crucial at THz frequencies). However, the THz multi-user channel is very sparse, and most of the analytical frameworks must be revised in the THz context. THz RIS systems can also assist virtual reality applications, as illustrated in \cite{chaccour2020risk}.

Several attempts for channel modeling in IRS-assisted scenarios at high frequencies are noted \cite{boulogeorgos2021pathloss}. In \cite{Mehrotra8823296}, a 3D channel model for indoor hypersurface-assisted communications at $\unit[60]{GHz}$ is developed. Channel estimation for IRS-assisted THz communications is studied along with hybrid beamforming in \cite{ning2019channel}, where cooperative channel estimation is achieved via beam training and exploiting the advantages of high dimensionalities and poor scattering at THz frequencies. End-to-end 3D channel modeling of radiation patterns from graphene-based reflectarrays at true THz frequencies is also presented in \cite{Han7579223}. For an IRS consisting of $M_i\times N_i$ reflecting elements, the IRS-assisted NLoS communications system model by assuming AoSAs being placed at the transmitter and the receiver, is a simple extension of \eqref{eq:sysmodel} and can be expressed as \vspace{2mm}
\begin{equation}\label{eq:IRS1}
\mathbf{y} = \mbf{W}^{\Hr}_r \left(\mathbf{H}_{\rm{IR}}\mathbf{\Phi}\mathbf{H}_{\rm{TI}} \right)\mbf{W}^{\Hr}_t  \mathbf{x} + \tilde{\mathbf{n}},
\end{equation}
where $\mathbf{H}_{\rm{IR}}\in \mathbb{C}^{M_r N_r \times M_i N_i}$ is the channel between the IRS and the receiving array, $\mathbf{H}_{\rm{TI}}\in \mathbb{C}^{M_i N_i\times M_t N_t}$ is the channel between the transmitting array and the IRS, $\tilde{\mathbf{n}}$ is the equivalent noise vector at the receiver, and $\mathbf{\Phi} =  \hspace{.5mm} \text{diag} \left(\beta_{11} e^{j\Omega_{11}},\ldots,\beta_{M_i N_i}e^{j\Omega_{M_i N_i}} \right) \in \mathbb{C}^{M_i N_i\times M_i N_i}$ is the diagonal matrix that comprises the gains ($\beta$'s) and phase shifts ($\Omega$'s) at each IRS element.

Unlike at lower frequencies, where channel hardening effects can arise with large IRSs \cite{Jung8948323}, the channel is highly correlated and low-rank at high frequencies. Hence, in addition to increasing the signal strength, IRSs operating at high frequencies should enhance the system performance by increasing the overall channel rank and suppressing interference. The researchers in \cite{ozdogan2020using} demonstrate how an IRS can be used to increase the channel rank, leading to substantial capacity gains. 

{We illustrate this issue at THz} frequencies by extending the concept of spatial tuning in Sec. \ref{sec:SPtuning2} to IRS-assisted THz NLoS environments. In a simplified proof-of-concept binary IRS operation, we assume that each reflecting element can either fully absorb an incident signal or reflect it toward a target direction. Hence, by controlling which element to reflect, a spatial degree of freedom is added at the IRS level, which could enhance the multiplexing gain of the NLoS system, but at the expense of a reduced total reflected power. 
{Global solutions can be derived by} jointly optimizing $\Delta_1$, $\Delta_2$, and $\Delta_3$, the inter-SA spacing at the transmitting array, inter-reflect-element spacing at the intermediate IRS, and the inter-SA spacing at the receiving array, respectively, as illustrated in Fig. \ref{f:IRS}. For this special case, the channel matrices (less the molecular absorption factor) can be approximated as \vspace{2mm}
\begin{equation}\label{eq:IRS2}
\mathbf{H}_{\rm{IR}} \! \approx \! \frac{c}{4\pi f d_2}e^{-\frac{j2\pi f d_2}{c}}\! \begin{pmatrix}
  \omega_{1,1} & \omega_{1,2} & \cdots & \omega_{1,M_i N_i} \\
  \omega_{2,1} & \omega_{2,2} & \cdots & \omega_{2,M_i N_i} \\
  \vdots  & \vdots  & \ddots & \vdots  \\
  \omega_{M_r N_r,1} & \omega_{M_r N_r,2} & \cdots & \omega_{M_r N_r,M_i N_i}
 \end{pmatrix}\!\vspace{4mm}
\end{equation}
\begin{equation}\label{eq:IRS3}
\mathbf{H}_{\rm{TI}} \! \approx \! \frac{c}{4\pi f d_1}e^{-\frac{j2\pi f d_1}{c}}\! \begin{pmatrix}
  \bar{\omega}_{1,1} & \bar{\omega}_{1,2} & \cdots & \bar{\omega}_{1,M_t N_t} \\
  \bar{\omega}_{2,1} & \bar{\omega}_{2,2} & \cdots & \bar{\omega}_{2,M_t N_t}\\
  \vdots  & \vdots  & \ddots & \vdots  \\
  \bar{\omega}_{M_i N_i,1} & \bar{\omega}_{M_i N_i,2} & \cdots & \bar{\omega}_{M_i N_i,M_t N_t}
 \end{pmatrix}\! \vspace{2mm}
\end{equation}
where $\omega_{m_r n_r,m_i n_i} \!=\!  e^{-j\Psi_1(\Delta_2,\Delta_3)/2 d_2}$, $\bar{\omega}_{m_i n_i,m_t n_t} \!=\!  e^{-j\Psi_2(\Delta_2,\Delta_1)/2 d_1}$, $\Psi_1$ and $\Psi_2$ are functions of the specific geometry and coordinate system, and $d_1$ and $d_2$ are the distances between the centers of the IRS and the transmitting array and the IRS and the receiving array, respectively. 


Compressive sensing and machine learning techniques for IRS-assisted THz communications are promising. For instance, in \cite{taha2019enabling}, the training overhead for channel estimation and the baseband hardware complexity are both reduced by assuming a sparse channel sensor configuration for surfaces. In this architecture, several elements in the IRS remain active (without RF resources; standard reflecting elements cannot send pilot symbols for channel estimation), and compressive sensing is used to acquire the channel responses on all other passive elements. This knowledge can then be exploited in a deep learning-based solution to design the reflection matrices with no training overhead. For IRS systems with imperfect CSI, distributed reinforcement learning techniques are also considered for channel estimation in \cite{zhang2020millimeter}. Deep learning is also exploited for beam and handoff prediction in drones with IRSs in \cite{abuzainab2021deep}. Furthermore, deep reinforcement learning for IRS-based hybrid beamforming is studied in \cite{huang2021multi,huang2020hybrid}

Several other recent signal processing solutions for THz-IRS systems have been proposed. Sum rate maximization for IRS-assisted assisted THz communications is studied in \cite{pan2020sum,Hao9371019}. In \cite{ma2020intelligent}, indoor IRS-assisted THz communications are studied, where a near-optimal low-complexity phase shift search scheme is proposed as an alternative to a complex, exhaustive search. Beamforming in THz scenarios incorporating both graphene-based UM-MIMO arrays and metasurfaces is also studied in \cite{Nie9053786}, demonstrating the potential of combining these two technologies. IRS-based index modulation schemes \cite{Basar8981888} can also be very efficient at THz frequencies. A Taylor-expansion-aided gradient descent scheme is proposed in \cite{Chen9124799} for optimizing the desired phase shift combination of IRS elements. Cooperative beam training using hierarchical codebooks for IRS channel estimation is also proposed in \cite{Ning9325920}, and channel estimation for THz-IRS-MIMO systems is studied in \cite{Ma9145343}.

\section{Extensions}
\label{sec:extensions}


\subsection{THz Localization}
\label{sec:localization}

Instead of being a byproduct of the communications system, localization in 6G is indispensable for location-aware communications. High-resolution localization capabilities \cite{Rappaport8732419,Kanhere9356512,aladsani2019,kanhere2018,Absi2018,Shree2018,Absi8454696,wang2019methods,Stratidakis8885710,kanhere2021outdoor} are critical to THz communications, especially because the beams are narrow and mobile users are hard to track. Higher directionality, array compactness, and larger bandwidths are all features that can be exploited to enhance THz-based localization. LoS THz propagation conditions can significantly improve the required distance estimation for localization.

{In contrast, with tiny device footprints} and dense networks, THz localization becomes a prerequisite for communications. For example, channel estimation, spatial multiplexing, beam-forming, and resource allocation can benefit from THz-based location information. Furthermore, maintaining the relative positions of UEs is advantageous for efficient tracking and link re-establishment. THz localization thrives on the availability of adequate infrastructure and access to wider bandwidths, where cooperative localization can further improve the system performance. Therefore, the interaction between THz localization and communication can synergistically contribute to a versatile system that can perform multiple functions beyond data communications. 

The angular and triangular accuracy in 3D localization can be enhanced by THz-band massive array signal processing, ensuring accurately beamformed THz beams in the elevation and azimuth directions of interest. Furthermore, conventional ranging techniques based on the received signal strength, time of arrival, angle of arrival, and time difference of arrival can be adjusted for THz characteristics. Nevertheless, novel localization techniques are more suitable for high-frequency operations. For instance, simultaneous localization and mapping (SLAM) can use THz-generated high-resolution environment images to improve location precision \cite{aladsani2019}. {Multidimensional scaling (MDS) is another technique} that can be used for THz network localization \cite{sarieddeen2019generation}. 

THz localization algorithms can use both IRS phase shifting and base station precoding for the delay and angle estimation. THz radar also promises to achieve millimeter accuracy. In the context of IRS systems, by acquiring the accurate position of an UE, joint precoding at the base station and phase shifting at the IRS can guarantee accurate angle or delay estimation. In \cite{ghasempour2020single,ghasempour2020leakytrack}, a leaky-wave antenna with a broadband transmitter is proposed for single-shot link discovery of neighboring nodes and mobility tracking. Learning-based solutions for high-frequency localization are also promising \cite{zhang2020learning,burghal2020comprehensive,Fan9155458}. THz sensing, imaging, and localization applications can all be piggybacked onto THz wireless communication or supported via dedicated resource allocation schemes \cite{sarieddeen2019generation}.





\red{

}

\begin{figure*}
\begin{equation}\label{eq:sens}
\begin{aligned}
 \mbf{y} =  \underbrace{\begin{bmatrix}
     h_{1,1}e^{-\frac{1}{2} \sum_{g=1}^G\mathcal{K}^g (f_1) d_{1,1}   } & 0  & \cdots & 0\\
    0 & h_{2,2}e^{-\frac{1}{2} \sum_{g=1}^G\mathcal{K}^g (f_{2}) d_{2,2} } & \cdots & 0\\
    \vdots & \vdots  & \ddots & \vdots \\
    0 & 0 & \cdots & h_{M_rN_r,M_tN_t}e^{-\frac{1}{2} \sum_{g=1}^G\mathcal{K}^g (f_{M_tN_t}) d_{M_rN_r,M_tN_t} }\\
   \end{bmatrix}}_{\mbf{H}}   \underbrace{\begin{bmatrix} x_1 \\ x_2 \\ \vdots \\ x_{M_tN_t}  \end{bmatrix}}_{\mbf{x}} + \mbf{n}.
\end{aligned}
\end{equation}   
\end{figure*}


\subsection{THz Sensing and Imaging}
\label{sec:sensing}

The unique THz spectral fingerprints of biological and chemical materials have been exploited in many sensing and imaging applications \cite{saeedkia2013handbook,Kianoush8682165,Woolard1512493,4337845Liu,fitch2004terahertz,Rappaport8732419,sarieddeen2019generation}, such as quality control, food safety, and security. Because THz frequencies are close to the optical realm, high-energy electromagnetic waves behave as photons, often interacting with other particles and matter. Such light-matter interactions with small particles (reflections, diffraction, and absorption) create unique electromagnetic signatures that can be exploited for THz sensing. 

THz signals can penetrate several materials and are strongly reflected by metals. They can also be used to analyze water dynamics (due to molecular coupling with hydrogen-bonded networks) and gas compositions (rotational spectroscopy). However, based on recent advances in THz technology and the prospect of realizing THz capabilities in hand-held devices, THz sensing applications will extend beyond the traditional industrial and pharmaceutical domains to reach everyday applications. Sensing, imaging, and localization applications will likely be piggybacked onto THz wireless communications \cite{sarieddeen2019generation}.

Novel THz-specific signal processing techniques are required to enable efficient joint THz sensing and communications. The first stage of THz sensing is signal acquisition, usually achieved via THz time-domain spectroscopy (THz-TDS). THz-TDS can be achieved in reflection mode or transmission mode, with the latter being more useful for sensing andimaging from a distance. Furthermore, reflection-based spectroscopy is more convenient in the context of joint communications and sensing. 

{Beyond signal acquisition,} several signal processing and machine learning techniques can be used to pre-process the received signals, extract characteristic features, and classify target materials into appropriate classes \cite{helal2021signal}. Furthermore, the accuracy of sensing and imaging is greatly enhanced in the THz band due to the vastly wider available channel bandwidths and the high directionality that accompanies massive MIMO beamforming. Smart metasurfaces operating in the THz band are also capable of sensing environments \cite{ma2019smart}. In \cite{Tasolamprou8788546}, artificial intelligence is used in the context of IRS-assisted intercell mmWave communications for sensing, programmable computing, and actuation facilities within each unit cell. 

{For massive THz MIMO systems}, carrier-based sensing and imaging can be more efficient as multiple RF chains can be tuned to multiple frequencies, generating multiple responses over the THz spectrum. Compared to short pulses covering the entire THz frequency range, carrier-based THz systems (frequency-domain spectroscopy) provide greater flexibility for choosing the carriers of interest for specific sensing applications. Only several carefully selected carriers can provide an efficient test for the existence of a specific molecule. 

In the particular case of carrier-based THz-band wireless gas sensing (also known as electronic smelling \cite{Kenneth8808165}), the estimated channel responses can be correlated with the HITRAN database \cite{gordon2017hitran2016} so that a decision is made on the gaseous constituents of the medium. We demonstrate that this sensing procedure can be seamlessly piggybacked over a communications system by considering an UM-MIMO AoSA scenario and assuming each SA to be tuned to a specific frequency symmetrically at the transmitter and receiver. Multiple SAs can still be tuned to the same frequency while assuming the channel to be orthogonal by design (following spatial tuning). Hence, the corresponding channel is diagonal, and the MIMO problem can be resolved into multiple single-input single-output problems. Each diagonal entry of $\mbf{H}$ thus represents the channel response between a particular SA at the transmitter (tuned to a particular frequency) and its corresponding SA at the receiver side. The received vector can then be expressed as in \eqref{eq:sens}, where $h_{m_rn_r,m_tn_t}$ is defined as \vspace{2mm}
\begin{equation}
\label{2}
    h_{m_rn_r,m_tn_t} = \mbf{a}_r^{\mathcal{H}}(\theta_r,\phi_r)G_r\alpha_{m_rn_r,m_tn_t}G_t\mbf{a}_t(\theta_t,\phi_t),
\end{equation}
\begin{equation} \vspace{2mm}
\alpha_{M_rN_r,M_tN_t}  =   \frac{c}{4\pi f_n d_{m_rn_r,m_tn_t}}  e^{  -j \frac{2\pi f_n}{c} d_{m_rn_r,m_tn_t}},
\end{equation}
with $\mathcal{K}^g (f_n)$ being the absorption coefficient of gas $g$ at frequency $f_n$. By solving for $\mbf{H}$, we identify both the gasses (or even specific isotopes of gases) that exist in the medium and their concentrations (by inspecting equation \eqref{abs_coeff}). The larger the AoSA size is, the more observations can be accumulated per channel use, and the faster the decision is made on the constituents of the medium.

Several methods can be used to solve for the absorption coefficients in $\mbf{H}$, including optimal maximum likelihood detectors, variations of compressed sensing techniques, and machine learning algorithms. For instance, instead of comparing the exact values of channel measurements, we can set thresholds to check the presence or absence of specific spikes and build decision trees for classification \cite{ryniec2012terahertz}. $\mbf{x}$ can be assumed to be a fixed or a random vector for sensing. However, in joint sensing and communications setups, the entries of $\mbf{x}$ would belong to a specific constellation with a specific structure. This knowledge can be exploited to enhance the sensing performance further.

\subsection{Networking and Security}
\label{networking_security}

\red{ 



}




Having addressed several THz-specific signal processing techniques, both the signal processing and networking problems differ in the THz band; both are linked to the underlying THz device architectures. Multiple access and networking paradigms for highly varying THz mobile environments are required. Hence, THz-specific MAC protocols \cite{ghafoor2019mac,yao2016tab,xia2019link,hossain2018terasim,Yao8849973,lemic2019survey,Zhang9316739,Cacciapuoti8387213,Polese9269934} need to be optimized jointly with PHY layer signal processing schemes under the constraints of state-of-the-art THz devices.  

{Examples of joint optimization schemes} include the studies in 
\cite{lemic2019assessing}, where energy harvesting THz nanonetworks are designed for controlling software-defined metamaterials, and in \cite{Yu9049790}, where joint THz power allocation and scheduling are optimized in mesh networks. Similarly, an on-demand multi-beam power allocation MAC protocol for THz MIMO networks is proposed in \cite{Yao8849973}. A receiver-initiated handshake procedure is proposed in \cite{xia2019link} to guarantee link-layer critical synchronization between high-speed THz networks. Synchronization of ultra-broadband THz signals, spectrum access and sharing, and neighbor discovery (given narrow beams) are open problems that require solutions on both the PHY and network layers.

THz security issues at the PHY and network layers are also important \cite{Sengupta9145177,singh2020thz,Yeh9296096,wang2021secrecy}. THz communications are more secure with higher propagation losses and increased directionality than communication paradigms at lower frequencies. Nevertheless, this enhanced security is not perfect. Security and eavesdropping in THz links are first studied in \cite{ma2018security}, where it is argued that security protocols should be designed on multiple levels, including hardware and the PHY layer. Signal processing techniques for waveform design can be proposed for the latter. 

{A scattering object can still} be placed within the broadcast sector of a transmitting antenna, despite the increased spatial resolution, which would then scatter radiation towards a nearby eavesdropper. By perfectly characterizing the backscatter of the channel, such a security breach can be avoided. Furthermore, narrow beams can still cover a relatively large area around the receiver, a vulnerability that can be exploited for eavesdropping. In \cite{Petrov8845312}, this vulnerability is mitigated by THz multipath propagation at the expense of slightly reduced capacity (sending shares of data over multiple paths). In other notable studies, secure-transmission IRS-assisted THz systems are studied in \cite{Qiao9120206}, and adding receiver artificial noise to enable THz secure communications is proposed in \cite{Gao9322487}.

Covert THz communications, in which an adversary residing inside the beam sector is prevented from knowing the occurrence of transmission, are also gaining attention. For instance, in \cite{Liu8964397}, covert THz communication is studied at the network level in dense internet-of-things systems using reflections and diffuse scattering from rough surfaces. Furthermore, in \cite{Gao9013380}, covertness is achieved by designing novel modulation schemes, such as distance-adaptive absorption peak hopping, in which frequency hopping is strategically selected at THz molecular absorption peaks. The covert distance is dictated by the transmit power and the SNR thresholds.





\section{Conclusions}
\label{sec:conc}


In this paper, we present a first-of-its-kind tutorial on signal processing techniques for THz communications. We detail the THz channel characteristics and summarize recent literature on THz channel modeling attempts, performance analysis frameworks, and experimental testbeds. We also highlight problem formulations that extend classical signal processing for wireless communications techniques into the THz realm. We study THz-band modulation and waveform design, beamforming and precoding, channel estimation, channel coding, and data detection, extending the discussion to the role of reflecting surfaces in the THz band and THz sensing, imaging, and localization. We also highlight THz-band networking and security issues. The techniques discussed in this paper will continue to evolve in the near future, driven by advances in THz transceiver design and system modeling.

\section{Acknowledgments}
\label{sec:Acknowledgments}

We thank Prof. Josep Miquel Jornet and Dr. Onur Sahin for the fruitful discussions on the topic and Mr. Simon Tarbouch for his input on THz channel modeling.


\red{






}

\section*{Biographies}
\footnotesize

\textbf{Hadi Sarieddeen} (S'13-M'18) received his B.E. degree (summa cum laude) in Computer and Communications Engineering from Notre Dame University-Louaize (NDU), Lebanon, in 2013, and his Ph.D. degree in Electrical and Computer Engineering from the American University of Beirut (AUB), Beirut, Lebanon, in 2018. He is currently a postdoctoral research fellow at KAUST. His research interests are in the areas of communication theory and signal processing for wireless communications.

\textbf{Mohamed-Slim Alouini} (S'94-M'98-SM'03-F'09) was born in Tunis, Tunisia. He received his Ph.D. degree in Electrical Engineering from Caltech, CA, USA, in 1998. He served as a faculty member at the University of Minnesota, Minneapolis, MN, USA, then in Texas A\&M University at Qatar, Doha, Qatar, before joining KAUST as a Professor of Electrical Engineering in 2009. His current research interests include the modeling, design, and performance analysis of wireless communications systems.

\textbf{Tareq Y. Al-Naffouri} (M'10-SM'18) Tareq Al-Naffouri received his B.S. degrees in Mathematics and Electrical Engineering (with first honors) from KFUPM, Saudi Arabia, his M.S. degree in Electrical Engineering from Georgia Tech, Atlanta, in 1998, and his Ph.D. degree in Electrical Engineering from Stanford University, Stanford, CA, in 2004. He is currently a Professor at the Electrical Engineering Department, KAUST. His research interests lie in the areas of sparse, adaptive, and statistical signal processing and their applications, localization, machine learning, and network information theory.

\end{document}